\documentclass[prb,aps,nofootinbib,amssymb,twocolumn,groupedaddress,10pt]{revtex4-2}
\usepackage{project_macros-public}
\usepackage{project_macros-private}
\usepackage{svg}
\crefname{appendix}{Appendix}{Appendices}
\crefname{equation}{Eq.}{Eqs.}
\crefname{figure}{Fig.}{Figs.}
\crefname{table}{Table}{Tables}
\crefname{section}{Section}{Sections}
\creflabelformat{appendix}{[#2#1#3]}

\makeatletter
\AtBeginDocument{\let\LS@rot\@undefined}
\makeatother

\makeatletter
\renewcommand\onecolumngrid{
\do@columngrid{one}{\@ne}%
\def\set@footnotewidth{\onecolumngrid}
\def\footnoterule{\kern-6pt\hrule width 1.5in\kern6pt}%
}

\allowdisplaybreaks

\begin{document}
\title{\titlePaper}
\paperAuthors
\let\oldaddcontentsline\addcontentsline

\newcommand\SP[1]{\textcolor{red}{[SP: #1]}}

\newcommand\SHS[1]{\textcolor{blue}{[SHS: #1]}}

\renewcommand\figureautorefname{Fig.}
\begin{abstract}
We investigate the claims by Park and Haldane [{\it Phys. Rev. B} {\bf 90}, 045123 (2014)] of an intrinsic protected value of the electric dipole moment at the physical edge of fractional quantum Hall (FQH) systems. Contrary to prevailing expectations, we find that the edge dipole takes the expected intrinsic value only in certain very special cases. We identify key limitations in earlier numerical studies and employ density matrix renormalization group (DMRG) methods to accurately compute the ground-state dipole.
We focus on three representative systems: the $\nu=1/3$-vacuum edge, the  $\nu=2/3$-vacuum edge, and the interface between Pfaffian and anti-Pfaffian phases.  
We find that the expected intrinsic dipole value occurs only at $\nu=1/3$,
whereas the other systems do not exhibit the claimed intrinsic value.   We give arguments based on composite fermions as to why hierarchy states should generally not have protected intrinsic dipoles. These results have important implications for the energetics and edge structure of FQH states.
\end{abstract}

\maketitle

\noindent {\textit{Introduction.}}--- Around a decade ago, Park and Haldane (P\&H) argued that the edge of a fractional quantum Hall (FQH) sample carries an electric dipole moment arising from the balance between viscous and electric forces \cite{HaldanePark}. They related this dipole directly to the Hall viscosity, a known topological quantity \cite{Avron_1995,Read_2011,Wen_shift,Read_2009,Bradlyn_2012,haldane2009hallviscosityintrinsicmetric}, and concluded that the edge dipole is therefore topologically protected. This proposal was unexpected, as it identified a new and apparently universal edge property in an otherwise well-studied system. Specifically,  P\&H conjectured that at an interface between a FQH liquid and the vacuum,
\begin{equation}
    \frac{p^{x}_{\PH}}{L_y}
    =\frac{\eta_H}{B},
    \label{eq:HP_equation}
\end{equation}
where $p^{x}/L_y$ is the guiding-center dipole moment in the $x$-direction per unit edge length in the $y$-direction, $B$ is the magnetic field, and $\eta_H$ is the guiding-center viscosity of the FQH liquid.  
From here on, $p^{x}_{\PH}$ denotes the dipole value predicted by Park and Haldane~\cite{HaldanePark}, as given in Eq.~\eqref{eq:HP_equation}.
For an interface between distinct FQH liquids, Eq.~\eqref{eq:HP_equation} generalizes to involve the difference in guiding-center viscosities, $\eta_H \to \Delta \eta_H$.
Importantly, this definition is expected to hold in reference to a specified background charge density (discussed further below, Eq.~\eqref{eq:K_def} and \eqref{eq:dipoledef}), that is necessary to remove ambiguities in the definition of the dipole moment. 

The purpose of our paper is to argue against the claims of P\&H.  We find that the edge dipole moment $p^x$ is given by the prediction $p^x_{\PH}$ in Eq.~\eqref{eq:HP_equation}  only for a specific subset of FQH states, and even in those cases only on making certain stringent assumptions on the nature of the confining potential and the inter-particle interactions.

To extract $p^x$ numerically, we work on a cylinder with  with axial direction along $x$,  a circumference $L_y$ and translation symmetry only along the circumferential direction $y$.  
We use Landau gauge, so each single-particle Landau-level orbital is labelled by an integer $j$ corresponding to its momentum along the $y$ direction as well as representing its average $x$-position, $x_j =2 \pi j \ell_B^2/L_y$. 
The quantum Hall Hamiltonian conserves particle number and is also translationally invariant along \(y\).    Consequently, both the total particle number  and the total momentum along \(y\) direction are good quantum numbers.  
We will relate the guiding-center dipole to the total $y$-momentum. 

We define the $y$-momentum $K$ as 
\begin{equation}\label{eq:K_def}
    K\equiv\sum^{\infty}_{j=-\infty} [n_j-\nu_0(j)]\, j,
\end{equation}
where $n_j$ is the occupancy of $j$-th orbital and $\nu_0(j)$ is a defined reference density at the position of orbital $j$. For the case of a vacuum-FQHE interface, $\nu_0(j)=\nu \Theta(j)$, a step function with bulk filling $\nu$. 
Similarly, for an interface between two different FQH states, if both bulk states have the same filling, the background density becomes uniform: $\nu_0(j) = \nu$.  
The total number of electrons is chosen to exactly match the integrated background density.    There is no contribution from the bulk region, where $n_j=\nu$; and hence an infinite or half-infinite geometry can lead to a well defined momentum $K$.

Since the $y$-momentum is linked to the particle position in the $x$-direction, $K$ of the ground state is related to the dipole moment of the edge.   In particular, the guiding center dipole moment of a system is 
\begin{equation}
\frac{p^x}{L_y} = e\frac{2\pi \ell^2_B}{ L_y^2}\, K(L_y)   \label{eq:dipoledef}
\end{equation}
for any $L_y$.  (In the thermodynamic limit P\&H define the dipole as an integral over $k_y$. Here we follow their protocol for finite circumference and convert the integral to a sum in Eq.~\eqref{eq:K_def}). 

The claim of P\&H is that if we consider an FQH system with an edge in its ground state (for a fixed edge confining potential, or possibly no edge confining potential), the large $L_y$ limit of the right hand side of Eq.~\eqref{eq:dipoledef} will approach $\eta_H/B$ (as in Eq.~\eqref{eq:HP_equation}).  Similarly, for an interface between two FQH states, it is claimed that the right hand side approaches $\Delta\eta_H/B$.    Using Eq.~\eqref{eq:HP_equation} and Eq.~\eqref{eq:dipoledef} we define the expected value $K_{\PH}(L_y) = (\eta_H/B) (L^2_y /(2 \pi e \ell^2_B ))$ to be the value of $K(L_y)$ in the ground state expected by the arguments of P\&H (and using $\Delta \eta$ in the case of an interface between two fluids).

It is worth noting that the {\it fractional} part of $K(L_y)$, $K_{\rm frac}(L_y) \equiv (K \mod 1)$ is {\it also} fixed in terms of the Hall viscosity, by considerations involving the conformal field theory (CFT) describing the FQH liquid \cite{Zaletel_2013,Zaletel_2015}.
This implies that given a FQH system, $K(L_y)$ of its edge can only ever change in integer units.
However, since we are interested in the large $L_y$ limit of the right hand side of Eq.~\eqref{eq:dipoledef}, we will not be interested in $K_{\rm frac}(L_y)$.

Previous DMRG studies on the cylinder \cite{DMRG,topo-dipole} observed a dipole consistent with the prediction of P\&H.
However, this is likely a somewhat fortuitous coincidence, since it elided two important subtleties.
First, those works did not examine the role of an external confining potential or of details of inter-particle interaction, even though varying these is crucial to establish that the dipole is protected --- in which case, it is expected to persist for a wide class of potentials and interactions.
Second, both prior studies were restricted to a single momentum ($K(L_y)$) sector (which is conserved by the Hamiltonian), even though no physical principle forbids the ground state from appearing in a different sector; hence, it is possible that the true ground state has a distinct $K(L_y)$ from those found by Refs.~\cite{DMRG,topo-dipole}, and hence deviate from the predicted `intrinsic' value. 
In this work, we investigate interfaces of different quantum Hall systems while explicitly incorporating these previously overlooked considerations.

In order to conclude that we have a protected dipole moment, the value of the moment  must be robust to (small) changes in details of the confining potential and details of the interaction.   Since $K(L_y)$ and hence the dipole moment $p^x$  (see Eq.~\eqref{eq:dipoledef}) is a constant of motion, we can consider the ground state energy of the Hamiltonian in each $p^x$-sector giving a function $E(p^x)$ which should have a minimum at the predicted value $p^x_{\PH}$.    This minimum should not be smooth, but rather should be a cusp (a discontinuity in $E(p^x)$ or in the slope $dE(p^x)/dp^x$)  since we can add a small linearly confining potential near the edge which  would convert $E(p^x) \rightarrow E(p^x) + \alpha (p^x - p^x_{\PH}) + \mbox{const}$, for some constant $\alpha$.  If the minimum were smooth, its location would shift depending on the confining potential.     Strictly speaking, we would want the minimum of $E(p^x)$ at $p^x_{\PH}$ to be a stable global minimum of the energy.  
However, one may also adopt a weaker criterion, whereby a protected dipole is said to exist if \(E(p^{x})\) exhibits appropriate non-analytic behaviour at \(p^{x} = p^{x}_{\PH} \), but is not a global minimum. 
In this situation, an appropriately engineered edge potential could, in principle, promote such a local minimum to a global one.

Our results modify the prevailing picture of dipole formation at quantum Hall interfaces.  We find that for general FQH states, the dipole moment need not assume the value predicted by P\&H, and can depend sensitively on system-specific details such as the interaction and the external potential. 

Specifically, we find that Laughlin states \cite{Lauglin_original_paper} at filling $\nu = 1/m$ obey the P\&H prediction in an appropriate regime, whereas several types of more complex FQH states do not.
In this work we give numerical results for several representative FQH states. The first system we consider is a $(\nu=\frac{1}{3})$/vacuum interface (i.e., the Laughlin edge), where we indeed find a  protected dipole as predicted by P\&H, under certain conditions.
On the other hand, for the $(\nu = \frac{2}{3})$/vacuum interface, we find that the ground state does not have a protected dipole value.
Near the $p^x=p^x_{\PH}$ sector, localized lone quasiholes are present and can move freely to change $p^x$ smoothly, without any abrupt energetic cost.  As a result, $p_{\PH}^x$ is not a particularly special value. 
The last system presented here is the Pfaffian/AntiPfaffian interface in the absence of any external potential. We show that in this case the ground-state dipole differs from $p^x=p^x_{\PH}$, with the system favouring a state with essentially no dipole. 
We have also investigated several other systems besides these; a full list is provided in the SI~\cite{Supplement}.

\noindent
{\textit{Model and Methods.}}--- We employ ``segment'' DMRG  \cite{DMRG_white,dmrg_scholl} in the cylinder geometry. Details of the DMRG methods for quantum Hall systems can be found in Refs.~\cite{Zaletel_2013,Supplement,Zaletel_2015,Zalatel_taige_newest,lotric_2025}; here we briefly summarize the key ingredients, with emphasis on aspects relevant to the interface construction used in this work. 
Tensor network simulations were performed using the TeNPy Library \cite{tenpy}.

In a typical FQH/vacuum interface, we consider $N$ electrons in $M$ orbitals labelled $j=0,1,\dots,M-1$. For orbitals $j<0$, we impose a hard wall, enforcing $n_j = 0$. The system size  $M$ is chosen so that the average filling in the segment matches the bulk filling, $ \nu_{\rm bulk}=N/M$. This procedure has been employed in previous studies~\cite{DMRG}.
Within this framework, we consider either a screened Coulomb interaction or interaction profiles specified by Haldane pseudopotentials $\{V_1, V_3, \dots\}$.

The generic construction of the interface proceeds via the cut-and-glue method described in Ref.~\cite{topo-dipole}, and applies to both the FQH/vacuum and FQH/FQH cases.
The central idea of segment DMRG is that the wavefunction is constrained to approach a prescribed bulk wavefunction on both the left- and right-hand sides through the use of matrix product state (MPS) tensors referred to as \emph{environments}.
The construction for the FQH/FQH and FQH/vacuum cases differs only in the choice of these environment tensors.
While the cut-and-glue approach enforces a fixed value of the momentum \(K(L_y)\), we develop a method that allows us to access and analyse different \(K(L_y)\) sectors. In previous works,~\cite{topo-dipole,DMRG} the value of \(K(L_y)\) was fixed to correspond to the intrinsic dipole, and no other momentum sectors were examined. As a consequence, the dipole moment was imposed as a conserved quantity, rather than emerging as a ground state from the energetics of the system. The detailed procedure for creating the interface is presented in the SI~\cite{Supplement}.

\noindent
{\textit{$(\nu=\frac{1}{3})$/vacuum interface.}}--- Consider the edge of a $\nu=\frac{1}{3}$ system. In the case where the electrons interact via the $V_1$ Haldane pseudopotential, Laughlin's wavefunction is the exact ground state in the bulk. For this state, inserting a quasihole into the bulk incurs no energy cost, whereas inserting a quasielectron does (\cite{hierarchy_haldane,Pokrovsky_1985,TrugmanKivelson}).

\begin{figure}
    \centering
    \includegraphics[width=\linewidth]{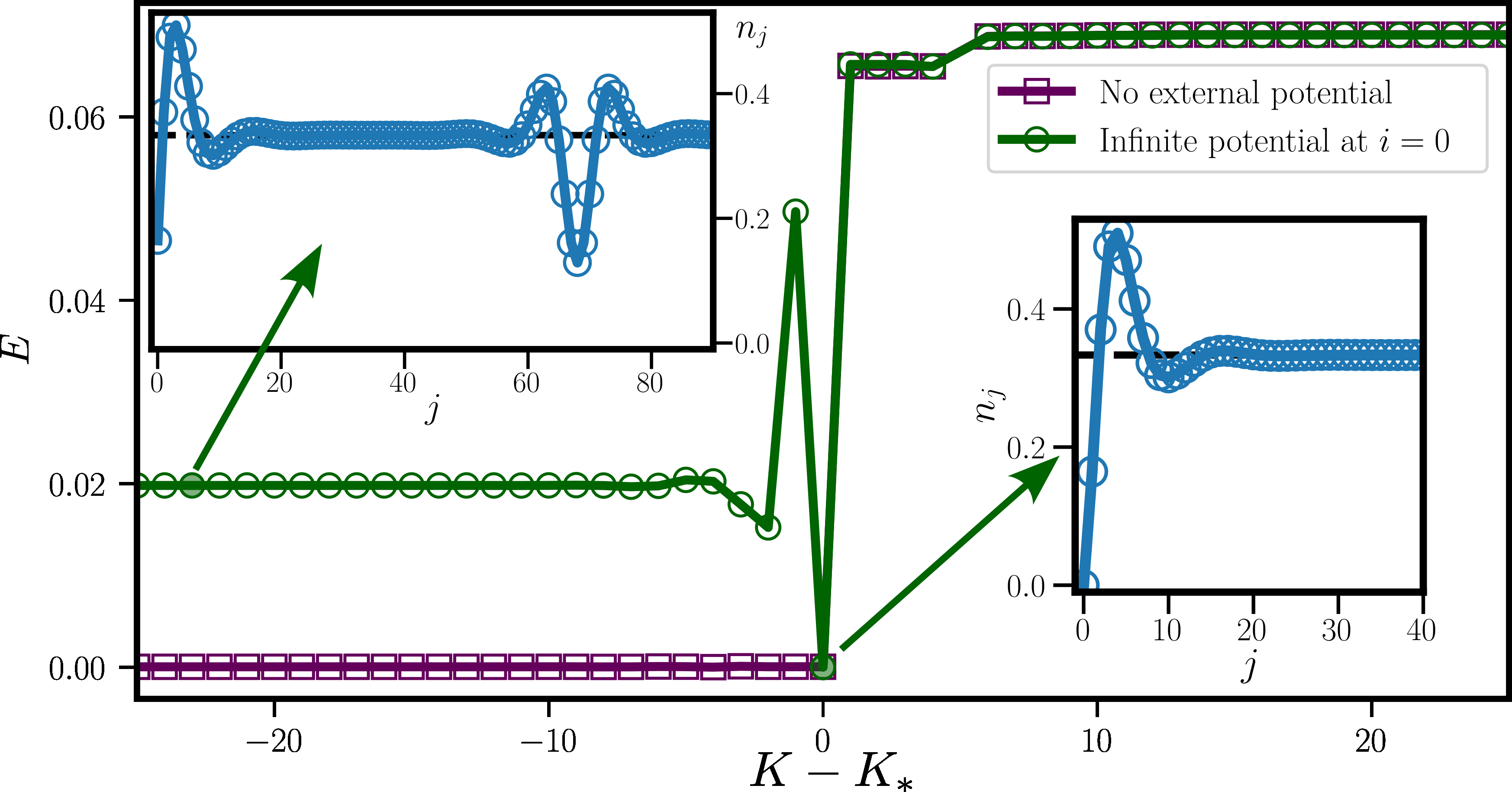}
    \caption{\textbf{($\nu=\frac{1}{3}$)/vacuum interface with $V_1$ pseudopotential interaction.} 
    The profile of $E$ vs. $K(L_y)$ for two different choices of external potential. (Inset)  Density profiles with the second choice, for $ K(L_y)=K_{\PH}(L_y) - 23$ (left)  and the P\&H dipole value ($ K(L_y)=K_{\PH} (L_y)$, right). The former case shows a quasihole deep in the bulk. Here $L_y=18 \ell_B$, and length of the segment is $M=167$.}
    \label{fig:DMRG:one third without V3}
\end{figure}

The simplest potential we consider is the hard wall discussed above, which forbids occupancy of orbitals $j<0$, while no potential is applied to orbitals with $j \geq 0$. We enforce the constraint that the average filling in the segment from $j=0, \ldots, M-1$ is exactly $\nu=1/3$.
The energy profile is shown in  Fig.~\ref{fig:DMRG:one third without V3}.
The Laughlin wavefunction satisfies $K(L_y)=K_{\PH}(L_y)$ and therefore has the dipole moment predicted by P\&H.    
A simple inspection of the Laughlin ground state wavefunction shows that the orbital at position 
$j=0$ is unoccupied. 
This follows directly from counting powers in the polynomial expansion of the wavefunction, and can also be seen numerically in  Fig.~\ref{fig:DMRG:one third without V3}.    
For this particular Hamiltonian the ground state in the  $K(L_y)=K_{\PH}(L_y)$ sector has the same energy as the ground state energy for any sector $K(L_y)<K_{\PH}(L_y)$.  
The reason for this is that for this special interaction a quasihole costs no energy. 
Starting with the Laughlin ground state, one can make a quasihole in the bulk at any orbital $j > 0$.  The density that is pushed away from the hole can be fully accommodated in the (previously empty) $j=0$ orbital without any energy cost. 
In the inset of Fig.~\ref{fig:DMRG:one third without V3} one can see the quasihole deep in the bulk. 

Because the ground state is degenerate for any $K(L_y) < K_{\PH}(L_y)$, strictly speaking this is not a stable dipole.
However, since we can add a weak confining potential to increase the energy of all $K(L_y) < K_{\PH}(L_y)$,  and since $K(L_y)>K_{\PH}(L_y)$ requires introducing a quasielectron which is gapped, we conclude that the ground state at $K(L_y)=K_{\PH}(L_y)$ should be stable for a range of confining potentials and we confirm the P\&H dipole in this case. 

Another example of a potential stabilizing a P\&H dipole is given by adding an infinite on-site potential introduced at orbital $j=0$. This deformation does not change the wavefunction in the Laughlin (ground state) sector.  
The resulting ground state sector coincides with the $K(L_y)= K_{\PH}(L_y)$ and the energy  clearly has a sharp minimum at this value (as shown in Fig.~\ref{fig:DMRG:one third without V3}).

To test the dependence on the interaction, we consider several interaction potentials beyond the pure $V_1$ case.
For screened Coulomb interactions with a hard wall at $j\leq 0$, the cusp at $K(L_y)=K_{\PH}(L_y)$ remains. For an interaction with $V_1 > 0$ and $V_3<0$, for small $|V_3|$ the the ground state remains in the $K_{\PH}(L_y)$ sector. Upon further making $V_3$ more negative, the ground state transitions to a different $K(L_y)$ value. However, $K_{\PH}(L_y)$ remains a local minimum of the energy so we cannot exclude the possibility that it would be stable under some range of confining edge potentials.
Detailed analysis is presented in the SI~\cite{Supplement}.

\begin{figure}
    \centering
    \includegraphics[width=\linewidth]{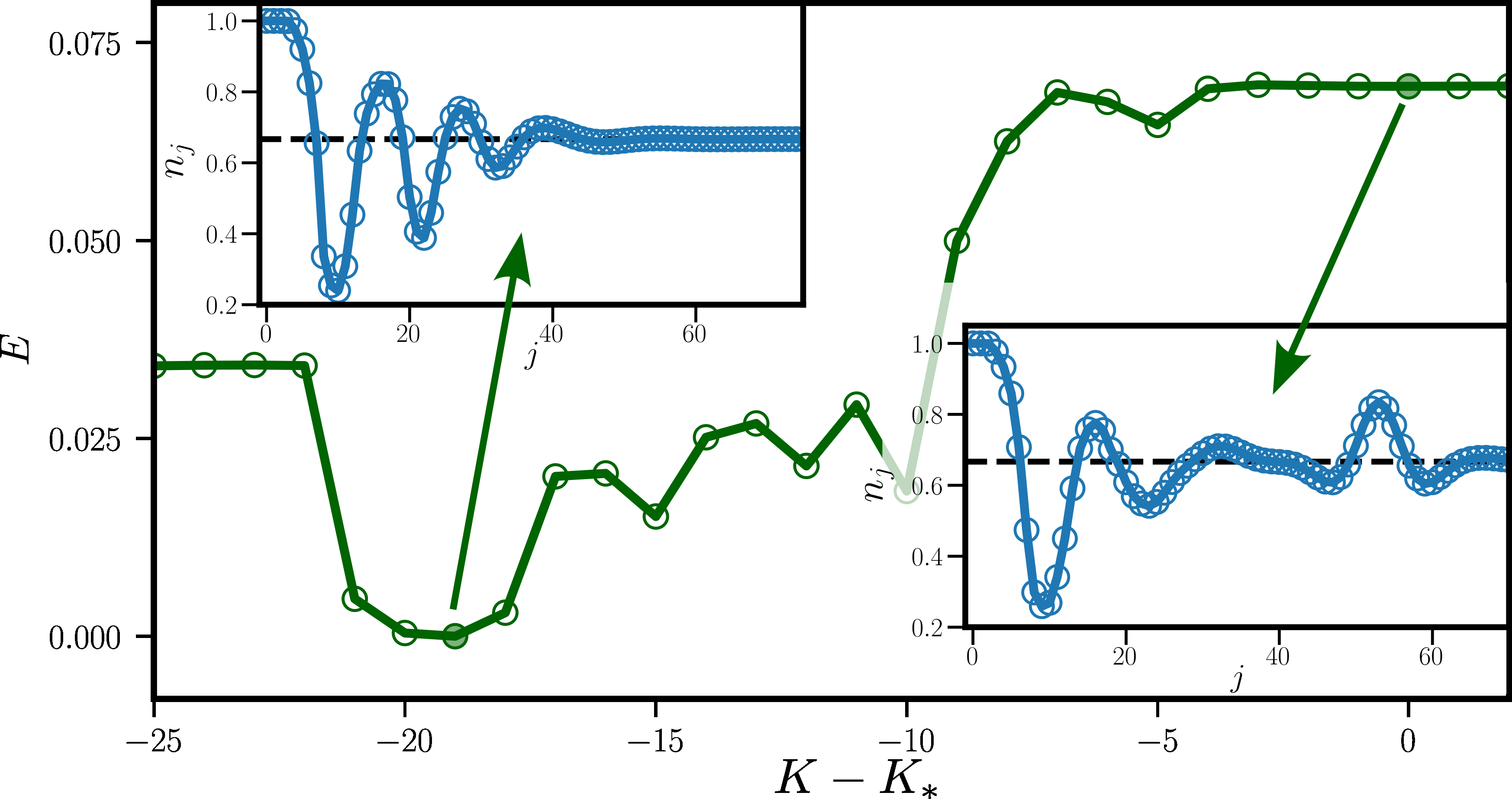}
    \caption{\textbf{$(\nu=\frac{2}{3}$)/vacuum interface with $V_1$ pseudopotential:} The profile of $E$ vs $K(L_y) - K_{\PH}(L_y)$ for a hard-wall potential. (Inset) Density profile  of the ground state sector (left) and the $K(L_y)=K_{\PH}(L_y)$ value sector (right). The circumference used for calculation is \(L_y=22 \ell_B\), but same qualitative behaviour was observed for \(L_y=12 - 25 \; \ell_B\). The length of the segment used was $M=185$.}   
    \label{fig:density_2_3}
\end{figure}

\noindent
{\textit{$(\nu=\frac{2}{3})$/vacuum interface.}}--- We now turn to the case of the \(\nu=\frac{2}{3}\) edge with a hard-wall confining potential, and a $V_1$ Haldane pseudopotential interaction,  analogous to the case we considered for $\nu=1/3$. 
The particle-hole 
conjugated Laughlin state is an exact ground state in the bulk and quasielectrons cost no energy. 
The energy as a function of $K(L_y)$ is shown in Fig.~\ref{fig:density_2_3} for $L_y= 22\ell_B$.  We find that the ground state sector is now $K(L_y)= K_{\PH}(L_y)-19$  and further, there is no cusp in the energy at $K(L_y)=K_{\PH}(L_y)$. 
As seen in Fig.~\ref{fig:density_2_3}, for $K(L_y)= K_{\PH}(L_y)$ there is a quasielectron deep in the bulk. 
Changing $K(L_y)$ then moves the quasielectron with no change in energy.   
The actual ground state occurs once this quasielectron merges with the edge. 
For $K(L_y) > K_{\PH}(L_y)$ the flat energy plateau extends arbitrarily far as we push the quasielectron further into the bulk.

Whereas the ground state does not occur at $K_{\PH}(L_y)$ for $L_y = 22\ell_B$ (and indeed, this might be considered as evidence enough against the arguments of P\&H), one might also wonder if the situation is different in the large $L_y$ limit.   
Since there are nontrivial finite-size effects, direct extrapolation of the minimum to large system size is not straightforward (although attempting such a direct extrapolation shows no sign that the ground state moves to $K_{\PH}(L_y)$, see Ref. \cite{Supplement}).
Instead, we aim to show that the minimum of $E(p^x)$ cannot occur at $p^x_{\PH}$ in the large system limit.  To do this, we consider fixed positive constant $a$ and consider $K_{\rm ref}(L_y) = K_{\PH}(L_y) - a L_y^2$ and show numerically that (a) $E(K)$ is monotonic between $K_{\PH}$ and $K_{\rm ref}$ and (b)  $E(K_{\PH}) - E(K_{\rm ref})$ is always positive and clearly non-decreasing as $L_y$ increases.
This then allows us to conclude that $p^x/L_y$ does not converge to $p^x_{\PH}/L_y$ in the large system limit.

\par
\begin{figure}[t!]
    \centering
    \includegraphics[width = \linewidth]{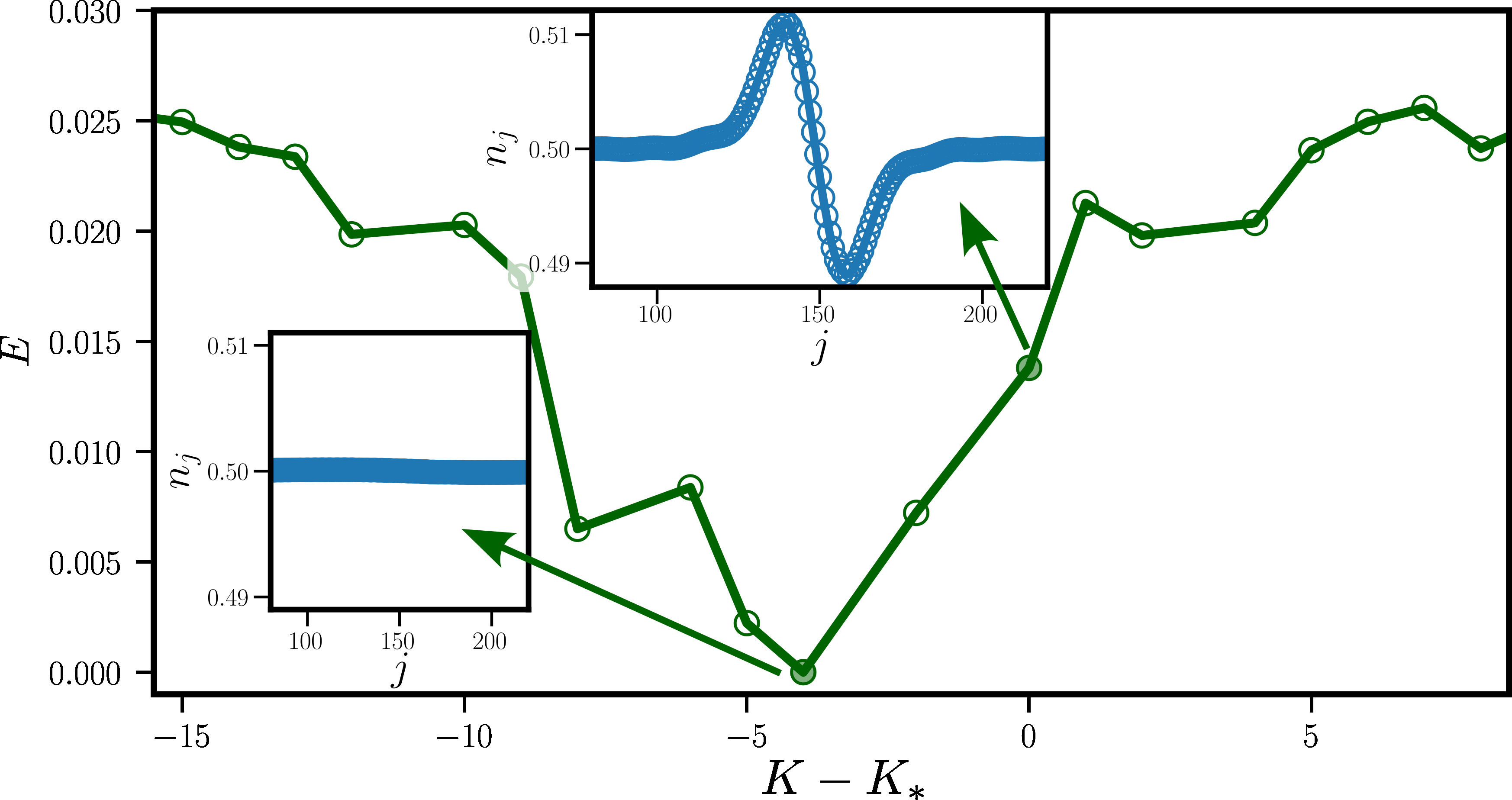}
    \caption{
    \textbf{Pfaffian/AntiPfaffian Interface:} The profile of $E$ vs $K(L_y) - K_{\PH}(L_y)$. Insets show the density profiles for the P\&H value of the dipole ($K(L_y)=K_{\PH}(L_y)$) and the ground-state dipole $K(L_y) -K_{\PH}(L_y)=-4$, which displays an almost flat density profile, so $p^{x}\approx 0.0$. Here $L_y=20 \ell_B$, $M=302$, and screening length is $\zeta=5.5\ell_B$.} 
    \label{fig:pf_apf} 
\end{figure}

\noindent {\textit{Pfaffian/AntiPfaffian interface.}}--- 
Next, we examine the Pfaffian–anti-Pfaffian (Pf–APf) interface for screened Coulomb interaction, in the absence of any external potential.
The corresponding energy profile $E(K)$ is shown in Fig.~\ref{fig:pf_apf}.  
It is evident that $K(L_y)=K_{\PH}(L_y)$ sector does not correspond to the global energy minimum. Instead, the minimum occurs near $K(L_y) =K_{\PH}(L_y)-4$, where the density profile is nearly flat (corresponding to $p^x \approx 0$). 
Importantly, the energy profile appears smooth and shows no clear cusp, suggesting that there is nothing special about $K(L_y)= K_{\PH}(L_y)$.  While finite size effects are hard to analyse for this system and larger systems are beyond our current numerical capabilities, we find similar results for all smaller systems, we find no sign of $K_{\PH}(L_y)$ being favoured, and the physical principle of minimizing dipole moment on the interface seems to always be obeyed. 
This is in direct contrast to previous results in the literature~\cite{topo-dipole}, presumably because they did not explore all $K(L_y)$ sectors. The data for different system sizes $L_y$ and screening lengths is presented in the SI~\cite{Supplement}.

\noindent {\textit{Composite Fermion Hierarchy.}}--- Among the usual composite fermion hierarchy states \cite{hierarchy_haldane,hierarchy_halperin}, for the reasons elaborated below, we conjecture that only interfaces between a Laughlin state and the vacuum will exhibit the P\&H protected value of the edge dipole. 

\begin{figure}[t!]
    \centering
    \includegraphics[scale=0.19]{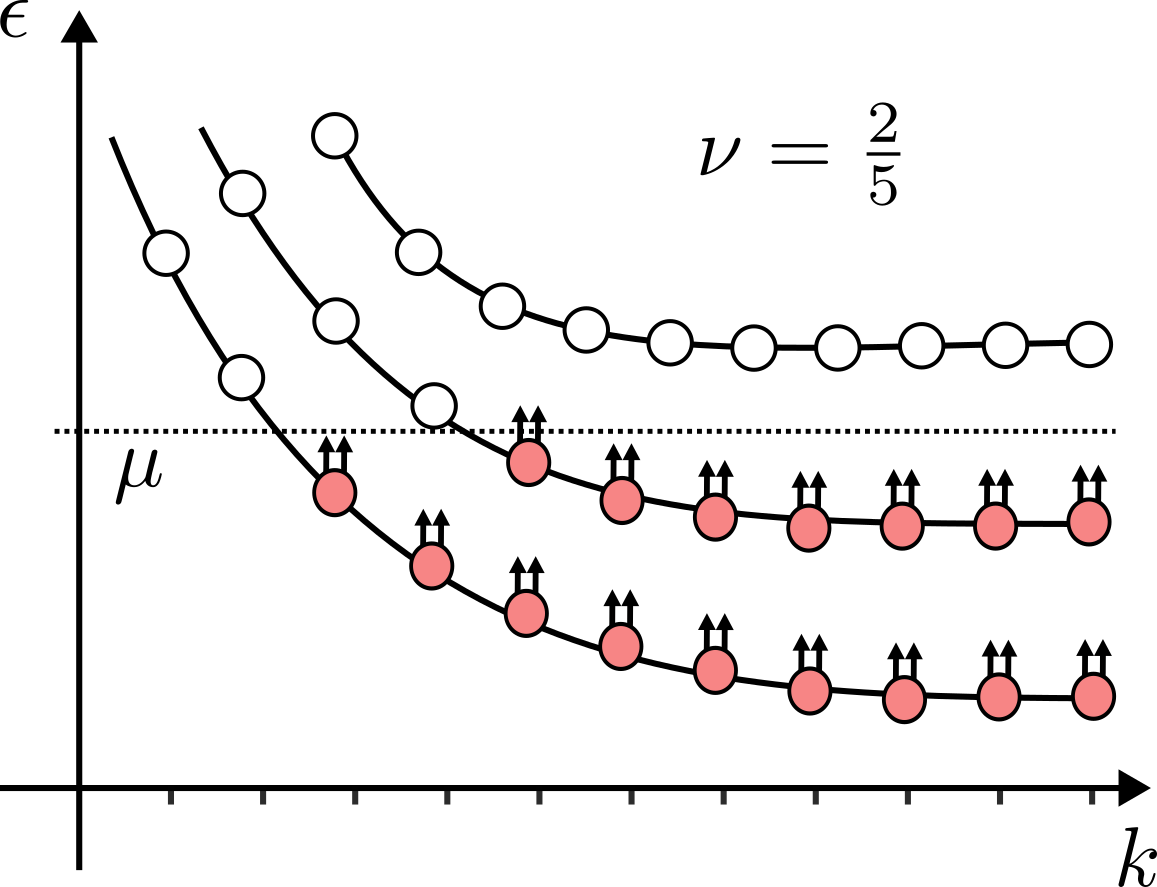}
    \caption{\textbf{Composite fermion dispersion:} The effective dispersion of composite fermion 'single-particle' wavefunctions within some confining potential. For a filling factor $\nu=\frac{2}{5}$, the dispersion exhibits a characteristic 'wedding-cake' structure, meaning that more composite fermions occupy the lowest $\Lambda$ level than the higher $\Lambda$ levels.}
    \label{fig:cf_cut}
\end{figure}

Jain states at filling $\nu = \nu^{*}/\left(2p\nu^{*}+1\right)$ with  integer $\nu^*$  can be understood as the composite fermions filling $\nu^{*}$ $\Lambda$-levels~\cite{Jain_CF_book,CF_original_jain,Lopez_fradkin}.   
We consider trial composite fermion wavefunctions on the cylinder for positive $\nu^*$, constructed using the Jain-Kamila projection~\cite{jain_kamila} adapted to the cylinder~\cite{pu_thesis,monte_carlo_topo_dipole}. 
By measuring their guiding-center dipole moments using Monte Carlo methods, we demonstrate that, for a composite fermion state to possess an `intrinsic' dipole, each $\Lambda$-level needs to be filled equally.
This means that all  $\Lambda$-levels are occupied up to the same momentum cutoff $k_y<|k_{\rm end}|$. 
Numerical data is shown for $\nu^* = 2,3$ ($p=1$) and $\nu^* = 2 $ ($p=2$) in the SI~\cite{Supplement}. 

However, near the edge one expects that all \(\Lambda\)-levels acquire an upward slope dependent on the edge confining potential.
The composite fermion states are then filled up to a common chemical potential $\mu$, and different $\Lambda$-levels generally intersect $\mu$ at different values of $k$ as illustrated in \autoref{fig:cf_cut} for $\nu=\frac{2}{5}$ \cite{Chklovskii_1995,Luis_Brey,jain_nu_2_5}.
As a result, we expect that the natural occupation has a “wedding-cake-like” structure, with the lowest $\Lambda$-level containing the largest number of composite fermions. 
Hence, we do not expect states with $\nu^{*}>1$ to attain an intrinsic dipole in the ground state.
The explicit numerics demonstrating this is quite challenging, so it is left for future work.
However, even for a hard-wall boundary for $\nu=\frac{2}{5},\frac{3}{7}$, where one might expect equal filling of the $\Lambda$-levels,  we still do not find the P\&H value of the edge dipole (see the SI~\cite{Supplement}).
For $\nu^{*}=1$, corresponding to Laughlin states, there is only one $\Lambda$-level so there is no “wedding-cake-like” structure and the P\&H dipole can be realized.

\par
\noindent {\textit{Summary.}}--- Our results contradict previous works: in particular, we show that the dipole moment on the boundary of a FQH fluid does not generically have to be stable or attain the value predicted by P\&H. However, there exists a class of FQH states, including the Laughlin states, for which the dipole does attain this value. We therefore conclude that the notion of an `intrinsic' dipole needs substantial further refinement, and at present the edge dipole moment cannot be viewed as a key  `protected' property of a QH edge, in contrast to other topological data inferred from the bulk-boundary correspondence.

\noindent {\textit{Acknowledgements.}}--- 
We thank Natalia Chepiga, Yuchi He and Tomohiro Soejima for valuable discussions. We thank Tomohiro Soejima for making suggestions on the Supplementary Information.
We are especially grateful to M. Zaletel, R. Mong, and F. Pollmann for granting us permission to use their code for compressing matrix product operators into MPO graph representations \cite{Zaletel_2013,Zaletel_2015}. We further acknowledge the use of the TeNPy library for tensor network simulations.
D.P. and K.V. acknowledge funding from Leverhulme Trust International Professorship Grant No. LIP-202-014.  S.A.P. and S.H.S. acknowledge support from EPSRC Grant No. EP/X030881/1.

\renewcommand{\addcontentsline}[3]{}
\bibliographystyle{apsrev4-2}
\bibliography{bibliography}
\let\addcontentsline\oldaddcontentsline

\renewcommand{\thetable}{S\arabic{table}}
\renewcommand{\thefigure}{S\arabic{figure}}
\renewcommand{\theequation}{S\arabic{section}.\arabic{equation}}
\onecolumngrid
\pagebreak
\thispagestyle{empty}
\newpage
\begin{center}
	\textbf{\large Supplementary Information for ``\titlePaper{}''}\\[.2cm]
\end{center}

\appendix
\renewcommand{\thesection}{\Roman{section}}
\tableofcontents
\let\oldaddcontentsline\addcontentsline
\newpage

\section{DMRG Methods}\label{app:DMRG_methods}
\subsection{MPS basics and notation}

\subsubsection{Schmidt decomposition and canonical form}
We briefly introduce the standard matrix product state (MPS) notation utilized in DMRG. Consider a one-dimensional system of length $L$, with local basis states $\{\ket{\sigma_j}\}$ at each site $j$. A generic many-body wavefunction can be written as
\begin{equation}
\ket{\Psi} = \sum_{\sigma_1,\dots,\sigma_L} \Psi_{\sigma_1 \dots \sigma_L} \ket{\sigma_1 \dots \sigma_L}.
\end{equation}
In the MPS representation, the coefficient tensor $\Psi_{\sigma_1 \dots \sigma_L}$ is factorized into a product of local tensors,
\begin{equation}
\Psi_{\sigma_1 \dots \sigma_L}
= \sum_{\{\alpha_i\}} B^{\sigma_1}_{\alpha_0 \alpha_1}
B^{\sigma_2}_{\alpha_1 \alpha_2}
\cdots
B^{\sigma_L}_{\alpha_{L-1} \alpha_L},
\end{equation}
where $\alpha_j$ are indices of auxiliary vector spaces living on the bond $\bar{j}$ between sites $j-1$ and $j$. 
We label bonds by $\bar{j}$ to distinguish them from sites.
The vector spaces have dimension at most $\chi_i$, and $\alpha_0,\alpha_L$ are dummy, boundary indices with $\chi_0=\chi_L=1$. Each tensor $B^{\sigma_j}_{\alpha_{j-1}\alpha_j}$ has one physical leg $\sigma_j$ and two auxiliary legs $\alpha_{j-1},\alpha_j$. This structure naturally arises from successive Schmidt decompositions of the wavefunction. For any bipartition of the system into left and right parts at bond $j$, the state can be written as
\begin{equation}
\ket{\Psi} = \sum_{\alpha_j=1}^{\chi_j} \lambda^{[j]}_{\alpha_j}
\ket{\alpha_j}_L \otimes \ket{\alpha_j}_R,
\end{equation}
where $\lambda^{[j]}_{\alpha_j}$ are the Schmidt coefficients, and $\{\ket{\alpha_j}_L\}$ and $\{\ket{\alpha_j}_R\}$ are orthonormal sets of $\chi_j$ states spanning the left and right Hilbert spaces $\mathcal{H}_L$ and $\mathcal{H}_R$, respectively. The states $\ket{\alpha_j}_{L,R}$ are referred to as the Schmidt states.
The auxiliary index $\alpha_j$ can thus be identified with the Schmidt states across the cut, and its dimension $\chi_j$ (the bond dimension) controls the amount of entanglement retained. 
There are different ways for how an MPS is constructed, due to the gauge-freedom $B^{\sigma_i}\rightarrow X_iB^{\sigma_i}X^{-1}_{i+1}$.
In one such gauge choice, the Vidal form, the MPS tensors are chosen such that the Schmidt decomposition is made explicit at each bond. One convenient representation is
\begin{equation}\label{eq:canonical_form}
B^{\sigma_j}_{\alpha_{j-1},\alpha_{j}}
= \Lambda^{[j]}_{\alpha_{j-1},\alpha_{j}} \Gamma^{\sigma_j}_{\alpha_{j-1},\alpha_{j}}
\end{equation}
where the diagonal (in auxiliary indices) matrices $\Lambda^{[j]}_{\alpha_{j-1},\alpha_{j}}$ contain the Schmidt values $\lambda^{[j]}_{\alpha_j}$, and the tensors $B^{\sigma_j}_{\alpha_{j-1},\alpha_j}$ are isometric. 
This representation is referred to as the Vidal canonical form.
This form makes it clear that the entanglement between two subsystems is fully characterized by the Schmidt coefficients associated with the bond $\bar{j}$ connecting them, while the $\Gamma$ tensors describe the change of basis between neighbouring Schmidt spaces.

\subsubsection{Conservation laws and charges}\label{sec:conservation_laws}

Consider local operator $\hat{q}_j$ acting on a single site in the DMRG, such that $\hat q=\sum_j \hat{q}_j $ commutes with Hamiltonian, and therefore $\langle\hat{q} \rangle$ is a conserved quantity.
The Schmidt states $\ket{\alpha_j}_{L/R}$ are obtained by diagonalizing the left and right reduced density matrices $\hat{\rho}_{L/R}$. The subsystem charge operators $\hat{Q}_{L/R}=\sum_{j\in L/R} \hat{q}_j$ are defined by summing the local charges over sites in the left and right subsystems, respectively. Because the reduced density matrices commute with these operators, $[\hat{\rho}_L,\hat{Q}_L]=0$ and $[\hat{\rho}_R,\hat{Q}_R]=0$, they are block-diagonal in the corresponding charge sectors.
Consequently, their eigenvectors can be chosen within sectors of fixed subsystem charge, so that each Schmidt state $\ket{\alpha_j}_L$ satisfies $\hat{Q}_L \ket{\alpha_j}_L = Q(\alpha_j)\ket{\alpha_j}_L.$
Similarly, each physical index $\sigma_j$ is associated with a charge $q(\sigma_j)$ determined by the local Hilbert space (e.g. particle number or momentum of the orbital).

\begin{figure}
\includegraphics[width=0.3\linewidth]{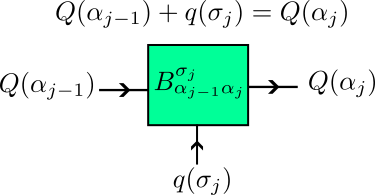}
 \caption{Figure representing the charge conservation in the MPS. The entry $B^{\sigma_j}_{\alpha_{j-1}\alpha_j}$ is non-zero only if $Q(\alpha_{j-1}) + q(\sigma_j) = Q(\alpha_j).$ }
     \label{fig:charge_conservation}
\end{figure}

Then, the MPS tensors are constrained by charge conservation, such that a tensor element $B^{\sigma_j}_{\alpha_j,\alpha_{j+1}}$ is non-zero only if the charges on its legs satisfy a fusion rule of the form
\begin{equation}\label{eq:conservation_charge}
Q(\alpha_{j-1}) + q(\sigma_j) = Q(\alpha_j).
\end{equation}
The fusion rule is illustrated in Fig.~\ref{fig:charge_conservation}.
As a result, the tensors decompose into independent blocks labelled by quantum numbers, and the MPS takes a block-sparse form. 
This structure significantly reduces computational cost and ensures that the variational state remains within a fixed symmetry sector.

We note that the value of the conserved quantity $\langle\hat{q} \rangle$ depends only on the physical charges $q(\sigma_j)$. 
Consequently, if all auxiliary charges are shifted uniformly as $Q(\alpha_j) \to Q(\alpha_j) + \Delta Q$, the conservation condition in Eq.~\eqref{eq:conservation_charge} remains satisfied, and the physical wavefunction is unchanged.
This reflects a gauge freedom in the MPS representation: while the auxiliary charge labels are modified, all physical observables and the state itself remain invariant.

\FloatBarrier
\subsection{Construction of the interface}
Here, we describe the construction of an interface between two distinct fractional quantum Hall phases, as well as between an FQH phase and the vacuum. The procedures for constructing FQH/FQH and FQH/vacuum interfaces are closely related, so we focus primarily on the FQH/FQH case and discuss the necessary modifications for the FQH/vacuum interface separately.
Our approach is based on the cut-and-glue construction used in Ref.~\cite{topo-dipole}, originally introduced by M. Zaletel, R. Mong, and F. Pollmann in Ref.~\cite{Zaletel_2013}.
We employ an extended version that enables systematic access to different momentum ($K$) sectors of the interface.
We note that the total number of electrons \( N \) in a segment of length \( M \) is fixed to match the bulk filling, i.e. \( \nu_{\rm bulk} = N/M \). 

The following provides an overview of our modified method for constructing an FQH/FQH interface consisting of a segment of \( M \) sites:
\begin{enumerate}
    \item We first obtain the ground states of the two distinct bulk phases independently using infinite DMRG (iDMRG), which directly targets the infinite cylinder limit ($L_x\rightarrow \infty,L_y$ finite), converging to a ground state with a unit cell of  $\tilde{M}$ sites.
    \item  
    We apply a gauge transformation that uniformly shifts the momentum quantum number $K$ on the auxiliary legs of the right bulk wavefunction by an amount $\Delta K$, while no shift is applied to the left bulk wavefunction. Since this is a gauge transformation, the bulk wavefunctions themselves remain unchanged; however, the way in which charges are assigned is modified. This enables systematic control over the total momentum sector of the resulting interface. 
    \item 
     We start from an infinite MPS. Truncating it at the right boundary yields a semi-infinite MPS. From this, we retain only a finite number \( \frac{M}{2} \) of sites near the truncation point, while all sites further to the left are absorbed into an environment tensor representing the bulk. This allows us to represent a semi-infinite cylinder using only \( \frac{M}{2} \) \( B \)-tensors together with a so-called environment tensor, which is fixed throughout.
     We apply this procedure to the other infinite MPS, but now truncate it at the left boundary and keep a $\frac{M}{2}$ sites on the right, with the remaining semi-infinite part encoded into the right environment.  
     We refer to this procedure as “cutting” the MPS.
     \item  The two truncated bulk states are then connected by contracting their auxiliary indices, with the bond structure enforcing matching conserved quantum numbers.
     Importantly, at the connected legs, the Schmidt values are replaced by random numbers.
     This state is in the correct symmetry sector, despite it potentially being a bad guess for the ground state.
    We refer to the process of joining the two truncated bulks into a single MPS as “gluing”.
    Changing the gauge on one MPS but not on the other essentially picks out different ways to contract auxiliary legs at the interface.
    \item Finally, we perform segment DMRG to optimize the finite region between the left and right environments, allowing the interface to relax to an optimized state.
\end{enumerate}

\begin{figure}
\includegraphics[width=0.6\linewidth]{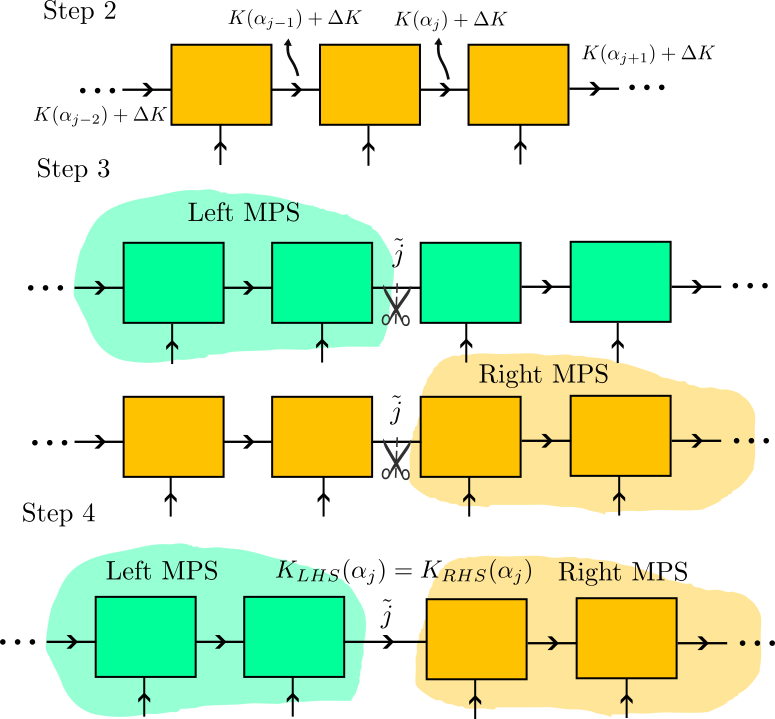}

 \caption{Figure illustrating steps 2--4 of the modified cut-and-glue procedure. In step 2, a uniform shift \( K(\alpha_j) \to K(\alpha_j) + \Delta K \) is applied to the auxiliary quantum numbers on all legs of one infinite MPS, while the other MPS remains unshifted.
 In step 3, the infinite MPSs are cut at bond $\tilde{j}$.
 Finally, in step 4, the two MPSs are glued together by matching the auxiliary charges at bond $\tilde{j}$, and discarding the legs that do not match.}
     \label{fig:cut_and_glue}
\end{figure}

Steps 2--4 are illustrated in \autoref{fig:cut_and_glue}.
The distinction between the standard cut-and-glue method and our modified approach is step 2, where we apply a gauge transformation to shift the momentum quantum number.
We describe this modification in detail in Sec.~\ref{subsec:charges}.
Furthermore, we provide additional numerical details on the construction of the FQH/FQH interface in Sec.~\ref{subsec:FQHE_FQHE}.
Finally, we generalize the construction to FQH/vacuum interfaces in sec.~\ref{subsec:FQHE_vacuum}, where we also present an alternative approach for constructing FQH/vacuum interfaces.

\FloatBarrier

\subsubsection{Variation of the momentum sector  \texorpdfstring{$K$}{K} }\label{subsec:charges}
In this subsection, we explain in detail how varying the momentum sector for the segment works. We consider the momentum operator $\hat{K} = \sum_j \hat{K}_j$, where $\hat{K}_j=j \hat n_j$ are local operators acting only on site $j$. The momentum operator $\hat{K}$ commutes with the Hamiltonian and is therefore a conserved quantity.

Now, consider the auxiliary bond $\tilde{j}$ (bond between sites $j-1$ and $j$). For the Vidal canonical form of an MPS (Eq.~\ref{eq:canonical_form}), the Schmidt decomposition at this bond is labelled by sectors $\alpha_j$ with Schmidt values $\lambda^{[j]}_{\alpha_j}$ and corresponding auxiliary charges $K(\alpha_j)$.
The expectation value of the local operator $\hat{K}_i$ can then be expressed in terms of the change in the weighted auxiliary charges across the bond:
\begin{equation}
\langle \hat{K}_j \rangle 
= \sum_{\alpha_{j+1}} \left(\lambda^{[j+1]}_{\alpha_{j+1}}\right)^2 K(\alpha_{j+1})
- \sum_{\alpha_j} \left(\lambda^{[j]}_{\alpha_j}\right)^2 K(\alpha_j).
\end{equation}

Similarly, the momentum on a segment $[i, j)$ is given by
\begin{equation}
\langle \hat{K}_{[i,j)}\rangle  \equiv \sum_{n=i}^{j-1} \langle \hat{K}_n \rangle 
= \sum_{\alpha_j} \left(\lambda^{[j]}_{\alpha_j}\right)^2 K(\alpha_j)
- \sum_{\alpha_i} \left(\lambda^{[i]}_{\alpha_i}\right)^2 K(\alpha_i).
\label{eq:supp:K segment}
\end{equation}

In the segment DMRG setup, we construct the segment wavefunction by fixing the left and right bulk wavefunctions and matching the auxiliary legs at the interface that carry the same conserved charges.
The total conserved momentum $\langle \hat{K} \rangle$ within the segment is therefore determined by the difference between the auxiliary charges of the right and left bulk wavefunctions.

As discussed in Sec.~\ref{sec:conservation_laws}, applying a uniform shift to all auxiliary charges does not change the physical wavefunction. This provides a way to modify the momentum of the segment without altering the bulk wavefunctions. 
Prior to constructing the interface, we apply the shift $K(\alpha_j) \to K(\alpha_j) + \Delta K$ to all auxiliary charges in the right bulk wavefunction. 
We cut the bulk wavefunctions by truncating them at the left (right) boundary, keeping a finite number of sites on the right (left), while the remaining semi-infinite part is absorbed into the right (left) environment.
We then connect the auxiliary legs of the cut left and right wavefunctions by matching the corresponding charges at the interface. As a result, the total momentum in the segment $0\leq j < M$ is shifted as
\begin{equation}
\langle \hat{K}_{[0,M)} \rangle \rightarrow \langle \hat{K}_{[0,M)} \rangle + \Delta K.
\end{equation}

Thus, by exploiting the gauge freedom prior to gluing the two bulk states, we can therefore select different conserved charge (or momentum $K$) sectors of the interface. 
In practice, shifting the auxiliary charges modifies which auxiliary legs can be consistently matched between the two bulk MPS representations; discarded legs correspond to incompatible quantum numbers, and different matchings realize different total charge sectors.

Finally, since the momentum sector of the segment is fully determined by the bulk wavefunctions, we can randomize the tensor entries within the segment without changing the total momentum sector. 
This allows us to generate a family of distinct initial trial wavefunctions that remain within the same conserved momentum sector.

 \subsubsection{Practical implementation for FQH/FQH interfaces}
 \label{subsec:FQHE_FQHE}

We note that the total conserved charge of the optimized segment is fixed entirely by the boundary conditions imposed by the left and right environments and is independent of the microscopic details within the segment. 
When gluing Pfaffian and anti-Pfaffian bulk states, their auxiliary charge spectra generally differ, and as a result some auxiliary legs must be discarded due to incompatible quantum numbers. By applying a uniform auxiliary charge shift prior to gluing, we effectively alter which subsets of auxiliary legs can be matched, thereby selecting different momentum (or conserved charge) sectors of the interface.

To generate distinct initial states for the segment DMRG optimization, we insert a small number of randomly initialized tensors between the cut Pfaffian and anti-Pfaffian wavefunctions.
While these tensors do not affect the total conserved charge of the segment, they modify the initial density profile and help avoid convergence to metastable local minima. 
In particular, directly matching Pfaffian and Anti-Pfaffian wavefunctions without inserting additional sites in between can result in a poor variational ansatz, as only a few auxiliary legs are compatible.
Introducing a small number of random sites improves the overlap of the auxiliary legs and can yield better initial states.
As an additional consistency check, we repeat the optimization with different numbers of sites in the central region and verify that all cases converge to the same energy and density profiles, confirming that the resulting interface state is independent of the specific initialization and segment length.

\subsubsection{FQH/vacuum interface}  \label{subsec:FQHE_vacuum}
The method described in Sec.~\ref{subsec:FQHE_FQHE} can also be applied to interfaces between the vacuum and an FQH phase. In this case, the right environment is given by the bulk FQH wavefunction, while the left environment corresponds to the vacuum, which can be represented by a $\chi=1$ bond dimension tensor. 
In addition, we impose a hard-wall boundary for $i<0$, ensuring that particles cannot propagate into that region.

There is also an alternative approach for constructing the FQH/vacuum interface:
On the vacuum side, we project the system onto the auxiliary leg with the largest Schmidt value in the charge neutrality particle number sector.
To vary the momentum $K$ sector, we append orbitals corresponding to either completely empty ($n_j = 0$) or fully occupied ($n_j = 1$) states. 
For example, in the $\nu = 1/3$ Laughlin state, a shift $\Delta K = 1$ is achieved by inserting the configuration $001$, while $\Delta K = 2$ corresponds to inserting $001001$, and so on.

The two approaches - charge-shifting of the auxiliary legs and orbital insertion - were benchmarked against each other for identical momentum sectors. 
Both methods produce consistent results for energy and density profile, confirming convergence and reliability.

\subsection{MPO construction}\label{sec:MPO_construction}
In MPO representation, any 2-body interaction is written as
\begin{equation}
V = \sum_{m_1,m_2,m_3,m_4} V_{m_1,m_2,m_3,m_4}\, c^\dagger_{m_1} c^\dagger_{m_2} c_{m_3} c_{m_4},
\end{equation}
where $V_{m_1,m_2,m_3,m_4}$ are the interaction matrix elements for orbitals given by momenta numbers $\{m_i\}$ in the Landau level basis. 
In constructing the MPO, we introduce a cutoff $\epsilon$ and discard all terms with $|V_{m_1,m_2,m_3,m_4}| < \epsilon$, typically taking $\epsilon = 10^{-5}$.

In this paper, we consider two types of interactions. 
The first is the Haldane pseudopotential interaction, specified by a finite set of coefficients $V_m$, which is short-ranged in the orbital basis and therefore admits an efficient MPO representation. 
The second is the Coulomb interaction.
A pure Coulomb interaction leads to matrix elements that decay slowly with orbital separation, resulting in a prohibitively large MPO bond dimension for any reasonable circumference $L_y$.
For this reason, we instead use a screened Coulomb interaction of the form
\begin{equation}
V(r) = \frac{1}{r}\, e^{-r^2/\zeta^2},
\end{equation}
where $\zeta$ is the screening length. We use different screening lengths for different systems, but typical range is $\zeta=2-6 \ell_B$.
This ensures sufficiently rapid decay of the matrix elements and allows for a controlled MPO construction. 
An efficient method for compressing MPO Hamiltonians of the FQHE systems into graph-like structures was developed by M. Zaletel, R. Mong, and F. Pollmann \cite{Zaletel_2013}. 
We use their algorithm to generate a compressed graph-MPO representation of the Hamiltonian.

\section{Jain composite fermion trial states and Variational Monte Carlo}
\label{app:CF_trial}
Motivated by the results presented in the main text, we now turn to an approximate approach to gain further insight into the DMRG data. In particular, we consider the system from a \textit{composite fermion} perspective. This framework provides access to a family of composite fermion trial states, with a natural way to construct distinct trial wavefunctions in different momentum sectors $K$. 
To this end, we employ variational Monte Carlo (VMC) to evaluate the variational energy of these trial states across different momentum sectors.
This allows us to compute $E(K)$, where $E$ denotes the variational energy.
For a given momentum sector $K$, the variational energy is defined as
\begin{equation}
    E(K) = \frac{\bra{\psi(K)} \hat{H} \ket{\psi(K)}}{\braket{\psi(K)}} ,
\end{equation}
where $\hat{H}$ is a two-body interaction Hamiltonian defined by an interaction potential $V(r)$, and $\ket{\psi(K)}$ denotes a trial ground-state wavefunction in the momentum sector $K$.

We choose \(\ket{\psi(K)}\) to be a Jain composite-fermion trial wave function. This choice is particularly convenient because the analytic form of the corresponding many-body wave function \(\Psi(z_1,\ldots,z_N)\) is explicitly known, which allows the expectation value of the Hamiltonian to be evaluated using standard variational Monte Carlo techniques. Within this framework, the variational energy can be expressed as an average of the interaction energy over the probability distribution defined by the wave function,
\begin{equation}
    E(K)=
\frac{
\int \prod_{i=1}^{N} \mathrm{d}^2 z_i \;
|\Psi(z_1,\ldots,z_N)|^2
\sum_{i<j} V(z_i-z_j)
}{
\int \prod_{i=1}^{N} \mathrm{d}^2 z_i \;
|\Psi(z_1,\ldots,z_N)|^2
} .
\end{equation}

In practice, the particle coordinates \(\{z_i\}\) are sampled from the probability distribution \(P(\{z_i\})=|\Psi(z_1,\ldots,z_N)|^2\) using the Metropolis algorithm. The interaction energy \(\sum_{i<j} V(z_i-z_j)\) is then averaged over these sampled configurations to obtain the variational energy. In particular, we apply this method to the \(\nu=\tfrac{1}{3}\) and \(\nu=\tfrac{2}{5}\) filling fractions, which serve as representative examples of Laughlin and Jain composite-fermion states. We now describe the details of composite fermion wavefunctions we use.

\subsection{Introduction to CF picture and Jain-Kamila projection}

The composite fermion (CF) approach provides a systematic framework for constructing accurate trial wave functions for fractional quantum Hall (FQH) states. In essence, electrons in a strong magnetic field are envisioned as binding an even number $2p$ of zeros of the wavefunction, which we roughly think of as flux quanta, to form composite fermions. These composite fermions experience a reduced effective magnetic field,
\begin{equation}
    B^{*}_{\mathrm{eff}} = B(1 - 2p\nu),
\end{equation}
where $B$ is the external magnetic field and $\nu$ is the electronic filling fraction. In this effective magnetic field, composite fermions fill $\nu^{*}$ effective Landau levels, known as $\Lambda$ levels. The corresponding electronic filling fraction is given by
\begin{equation}
    \nu = \frac{\nu^{*}}{2p\nu^{*} + 1}.
\end{equation}
For example, at $\nu = \tfrac{2}{5}$, composite fermions fill two $\Lambda$ levels. 
The usefulness of the composite fermion picture lies in its ability to map fractional quantum Hall states of electrons onto integer quantum Hall states of composite fermions.

We now describe how to construct a trial wave function for a Jain composite fermion state.
We work in a cylindrical geometry that is periodic along the $y$ direction with circumference $L_y$.
The single-particle states are labelled by the $\Lambda$-level index $n$ and an integer $k$ related to the momentum
\(
k_y = \frac{2\pi k}{L_y}.
\)
We first construct a Slater determinant corresponding to $n$ filled $\Lambda$ levels,
built from single-particle wave functions $\Phi^{(n)}_{k}(z_i)$.
Flux attachment is implemented by multiplying the Slater determinant by a Jastrow factor that attaches $2p$ vortices to each electron,
\begin{equation}
    \Psi_{\nu}^{\mathrm{CF}}(\{z_i\}) =
    \mathcal{P}_{\mathrm{LLL}}
    \left[
        \prod_{i<j} (w_i - w_j)^{2p}
        \;
        \det \Phi^{(n)}_{k}(z_i)
    \right],
\end{equation}
where $z_i = x_i + i y_i$ denotes the complex coordinate of the $i$-th electron,
\(
w_i = \exp\!\left( \frac{2\pi z_i}{L_y} \right),
\)
and $\mathcal{P}_{\mathrm{LLL}}$ denotes projection onto the lowest Landau level (LLL). The projection is essential, since for the states we are considering, physical electrons are restricted to the LLL.

However, exact LLL projection is exponentially costly in system size.
To circumvent this difficulty, we employ the Jain-Kamilla (JK) projection scheme~\cite{jain_kamila}, which provides an efficient approximate projection while preserving the essential physics of the composite fermion states.
Within this approach, the trial wave function is written as
\begin{equation}\label{eq:CF_equation}
    \Psi_{\mathrm{JK}}^{\mathrm{CF}}(\{z_i\}) =
    \prod_{i<j} (w_i - w_j)^{2p}
    \;
    \det \tilde{\Phi}^{(n)}_{k}(z_i),
\end{equation}
where $\tilde{\Phi}^{(n)}_{k}(z_i)$ are single-particle orbitals modified by the Jain-Kamilla projection to ensure compliance with the LLL constraint.
The explicit construction of these orbitals for the cylindrical geometry follows the prescription described in Ref.~\cite{pu_thesis}.
For completeness, we list below the explicit forms of $\Tilde \Phi^{(n)}_{k}(z)$ for the first three $\Lambda$ levels.

\begin{subequations}
    \begin{equation}
        \Tilde \Phi^{(0)}_{k}(z_i)=e^{-\frac{|z|^2+z^2}{4}} w^{k}_i
    \end{equation}
    \begin{equation}
       \Tilde \Phi^{(1)}_{k}(z_i)=e^{-\frac{|z|^2+z^2}{4}} w^{k}_i \left( \frac{2\pi k}{L}(\alpha-1)+\frac{2\pi p}{L}\sum_{j}\frac{w_i}{w_i-w_j}\right)
    \end{equation}
\begin{align}
\Tilde \Phi^{(2)}_{k}(z_i)
&= e^{-\frac{|z|^2 + z^2}{4}} \, w_i^{k} \Bigg[
\frac{\alpha(\alpha - 1)}{2}
+ \frac{4\pi^2 n^2}{L^2}(\alpha - 1)^2 \nonumber  
 + \left(\sum_j \frac{w_i}{w_i - w_j}\right)\frac{4\pi n \alpha (\alpha - 1)}{L} +\\&\quad 
  \alpha^2 \left(
\left(\sum_j \frac{w_i}{w_i - w_j}\right)^2
- \sum_j \frac{w_i^2}{(w_i - w_j)^2}
\right)\left(\frac{2\pi p}{L}\right)^2  
\Bigg].
\end{align}
\end{subequations}
with $\alpha=1-2p\nu$.

The construction of trial wavefunctions in different momentum sectors $K$ is based on the fact that the momenta of the composite fermion orbitals $k$ determine the total many-body momentum $K$.
By varying the occupation profiles of the composite fermions, we can therefore access different momentum sectors $K$.
In constructing the trial wavefunctions, several issues must be addressed. First, we must identify a state with the correct intrinsic dipole, corresponding to $K = K_{\PH}$. 
Such a state is constructed in sec. \ref{app:intrinsic dipole} and will be referred to as the ``reference'' state.
Next, we must account for the effects of the confining potential, as discussed in sec. \ref{app:boundary_condition}. 
With these ingredients in place, we can construct a family of trial wavefunctions across different momentum sectors, as detailed in sec. \ref{app:trial_CF_wavefunction_K_sectors}.

\subsection{Intrinsic dipole}\label{app:intrinsic dipole}

Here, we construct composite fermion states that possess an intrinsic dipole predicted by $P\&H$.
Generally, to describe a state with bulk filling fraction $\nu$, one must fill exactly $\nu^{*}$ $\Lambda$ levels in the bulk for each CF momentum number $k$. 
Near the edge, however, the occupation profile in the ground state may deviate from this uniform structure.
We define a “reference” state as one in which each allowed momentum $k_y$  is occupied exactly $\nu^{*}$ times all the way to the edge. 
If we have $N$ composite fermions, we use the convention that we fill up the orbitals with orbital momentum number $p \leq k\leq \frac{N}{\nu^{*}}+p-1$, where $2p$ is number of flux quanta attached to a wavefunction.\footnote{This choice ensures that the center of mass of the wavefunction in Eq. \eqref{eq:CF_equation} is consistent with the DMRG calculations, without requiring any additional translations. Equivalently, in the electronic language, the background density is given by $\nu_0(j)=\nu$ for $0 \le j < \nu N$, and $\nu_0(j)=0$ otherwise.}
We claim that such “reference” states satisfy $p^{x} = p_{\PH}$, and we verify this numerically.

\begin{figure}[ht]
\includegraphics[width=0.4\linewidth]{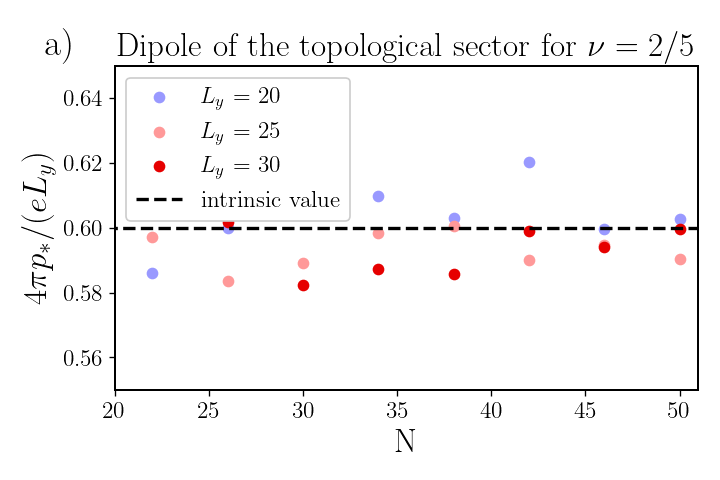}
\includegraphics[width=0.4\linewidth]{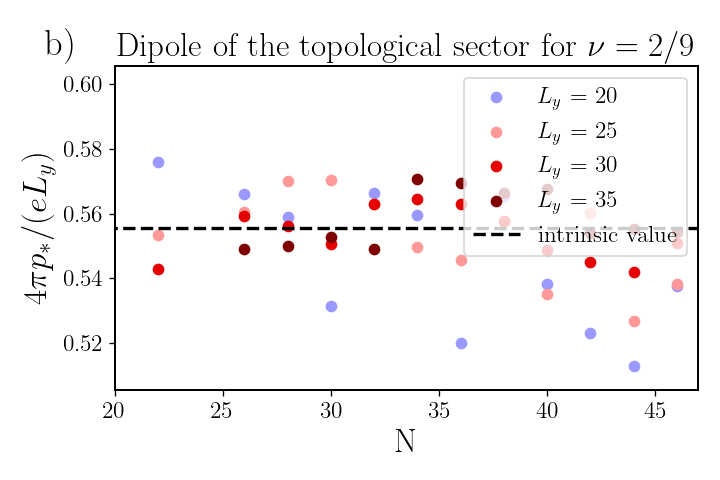}
\includegraphics[width=0.4\linewidth]{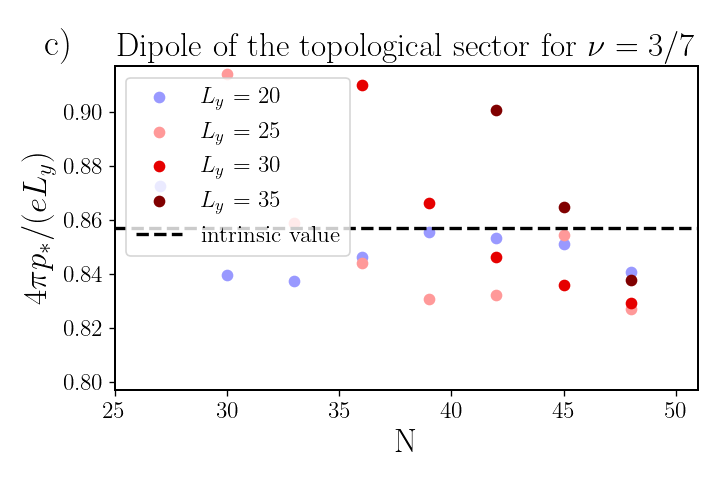}
 \caption{Subfigures a), b), and c) show the dipole for $\nu = \frac{2}{5}, \frac{2}{9}, \frac{3}{7}$, respectively, for the ``reference'' state, as a function of the number of composite fermions $N$ and the circumference $L_y$. These fillings correspond to $p = 1, 2, 1$ and $\nu^{*} = 2, 2, 3$, respectively. The dotted horizontal line indicates the P\& H dipole value. The overshoot occurs due to a small systematic error in measuring the center-of-mass position. We observe that these errors are comparable to the \( K \bmod 1 \) corrections to the dipole. }
     \label{fig:dipole_cf}
\end{figure}

We note that guiding-center dipole moment $p^{x}$ was previously computed for $\nu^{*} = 1$ with $p = 1,2$ (so $\nu=1/3,1/5$), and was found to agree with the intrinsic value of the dipole~\cite{monte_carlo_topo_dipole}.
We measure the real-space dipole moment defined by
\begin{equation}
    \frac{p^{x}}{L_y} = \int^{x_{\rm cut}}_{-\infty} x \left[\rho(x) - \rho_0(x)\right] \, dx,
\end{equation}
where 
\begin{equation}
    \rho_0(x) = \frac{2\pi \nu}{L_y} \sum_{j\geq 0} |\phi_{j,0}(x)|^2
\end{equation}
subtracts the contribution from the Landau orbital centers, with $\nu$ being the bulk filling, and $\phi_{j,0}(x)$ wavefunction in LLL with orbital number $j$.  
Here \( x_{\rm cut} \) is a cutoff chosen deep in the bulk, where \( \rho(x) = \rho_{0}(x) \); the bulk does not contribute to the dipole.
We calculate $p^{x}$ for states at filling $\nu = \tfrac{2}{5}, \tfrac{2}{9}, \tfrac{3}{7}$ and find that they attain an intrinsic dipole.
The numerical results are shown in \autoref{fig:dipole_cf}.

Furthermore, we observe that $K(L_y)$ can be recovered from the dipole $p^{x}$ via Eq. \eqref{eq:dipoledef} in the main text. In fact, by analysing data in \autoref{fig:dipole_cf}, we can make a stronger statement: $K(L_y) = K_{\PH}(L_y)$.
To see this, we note that a change of \( 1 \) in \( K (L_y)\) corresponds to a change of \( 0.064 \) in \( 4\pi p_{*}/(L_y e) \) at \( L_y = 35\, \ell_B \), so in all calculations in \autoref{fig:dipole_cf} we remain well within \( |K(L_y) - K_{\PH}(L_y)| < 1 \).

\subsection{Boundary conditions}\label{app:boundary_condition}
In order to proceed with the construction of trial composite fermion functions, we need to translate the DMRG confining potential from in electronic Landau level orbitals, to composite fermion $\Lambda$ level orbitals.
In particular, the hard-wall constraint imposed in the DMRG setup translates, within the composite-fermion picture, into a restriction on the allowed single-particle momenta within $\Lambda$ levels. 
In particular, if electrons are confined to occupy orbitals \(0 \leq j \leq N/\nu - 1\), then by counting powers of $w_i$ in Eq. \eqref{eq:CF_equation}, one finds that the orbital momentum of composite fermions must obey
\begin{equation}\label{eq:CF_constraint}
    0 \leq k \leq \frac{N}{\nu^{*}}+(2p-1)
\end{equation}
Here, \(k\) labels the composite-fermion orbitals, while \(j\) labels the usual Landau-level orbitals of electrons in the original magnetic field.  
Similarly, if the hard wall sets the region without the potential to \(0 < j \leq  N / \nu - 1\), this simply translates to $ 0 < k \leq \frac{N}{\nu^{*}}+(2p-1)$. 
Hence, the hard-wall potential constrains our composite fermion trial function basis.

\begin{figure}[ht]
\includegraphics[width=0.6\linewidth]{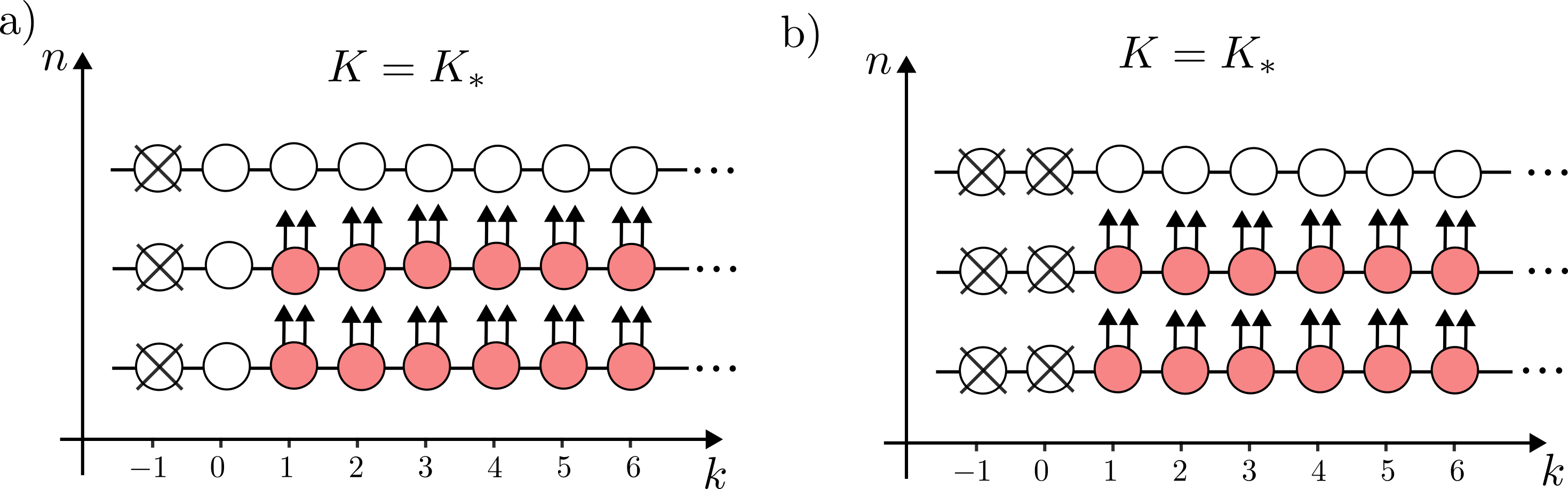}
         \caption{Figure representing the trial wavefunction occupation at $\nu=\frac{2}{5}$ ($\nu^{*}=2$, $p=1$) of composite fermions $\Lambda$ levels at $K=K_{\PH}$ sector with hard-wall confining potential at (a) $j<0$ and (b) $j\leq 0.$ As per our convention, the leftmost filled orbital has $k=p$.}
     \label{fig:cf_dispersion}
\end{figure}

Importantly, the ``reference'' state with $K = K_{\PH}$ occupies only $ N/\nu^{*} $ composite-fermion orbitals with distinct $k$ quantum numbers. 
Thus, if we use the hard wall potential that limits orbitals to $0\leq j\leq N/\nu-1$, there is $p$ lowest $\Lambda$ level unoccupied composite fermion orbitals at each side of the sample.
This is illustrated for the case of $\nu=\frac{2}{5}$ in \autoref{fig:cf_dispersion}.

\subsection{Different \texorpdfstring{$K$}{K} sectors within Jain-Kamila projection}\label{app:trial_CF_wavefunction_K_sectors}

Finally, we are ready to construct trial composite fermion wavefunctions for different confining potentials.
To construct trial states with $K \neq K_{\PH}$, we modify the occupancies near the edge. 
The basic idea is to promote a single composite fermion near the edge from a filled $\Lambda$ level to an empty orbital with a different orbital momentum $k$.
Changing $k$ alters the total momentum of the many-body state, allowing access to different momentum sectors $K$ within the same variational framework.
We note that promoting a composite fermion is equivalent to creating a quasielectron–quasihole pair.
While promoting a composite fermion in the bulk to a higher $\Lambda$ level is expected to cost a finite energy, near the edge such a promotion can, in principle, lower the energy relative to the “reference” state.

\begin{figure}[ht]
\includegraphics[width=0.9\linewidth]{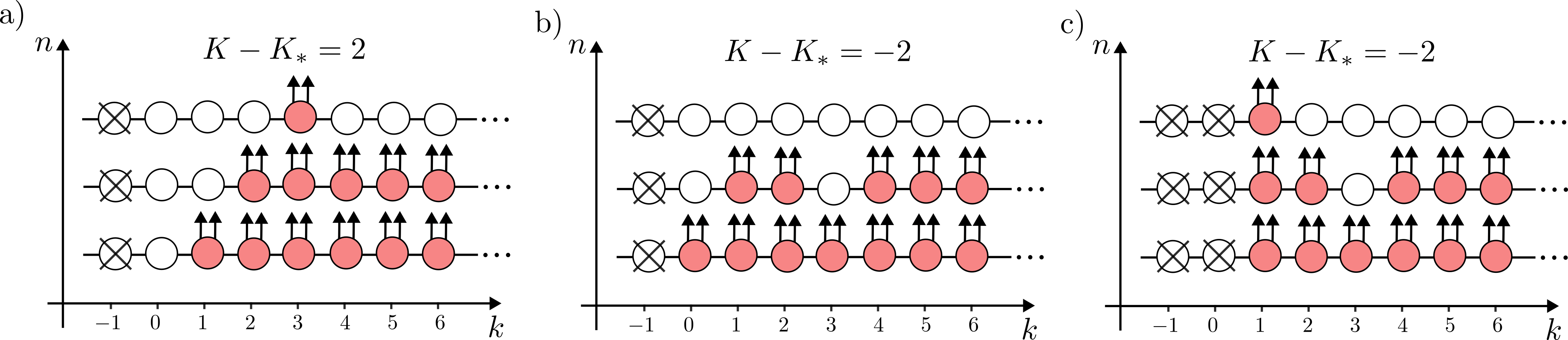}
         \caption{Figure representing trial wavefunction occupation of $\Lambda$ levels for different $K$ sectors for $\nu=\frac{2}{5}$ ($\nu^{*}=2,$ $p=1$). Subfigure a) represents $K-K_{\PH}=2$ sector with hard wall at $j<0$. Subfigure b) represents $K-K_{\PH}=-2$ sector with hard wall at $j<0$, whereas subfigure c) represents $K-K_{\PH}=2$ sector with hard wall at $j\leq 0$.}
     \label{fig:cf_dispersion2}
\end{figure}

In our particular construction, to obtain a state with $K = K_{\PH} + \Delta K$ and $\Delta K > 0$, we remove a single composite fermion from the “reference” state in the highest $\Lambda$ level at the leftmost orbital $k = 1$, and place it into an unoccupied orbital at $k = \Delta K+1$ in a higher $\Lambda$ level. 
Similarly, for $\Delta K < 0$, we remove a composite fermion from a highest occupied $\Lambda$ level at orbital $k = \Delta K + 1$ (or $k = \Delta K$), and move it to the leftmost unoccupied orbital at $k = 1$ (or $k = 0$), depending on whether a hard-wall constraint is imposed at $j < 0$ or $j \leq 0$. 
Here, we assumed $p = 1$, but the construction readily generalizes to other values of $p$.
Illustrations of such trial states are given in ~\autoref{fig:cf_dispersion2}.  We assume this construction throughout the paper.

Finally, we emphasize that this construction represents only a restricted subclass of variational states, as it involves promoting a single composite fermion.
In general, one can consider more elaborate variational wavefunctions involving multiple promoted composite fermions. We leave the study of such states for future work.

\FloatBarrier

\section{\texorpdfstring{$\nu=\frac{1}{3}$}{nu=1/3}/vacuum  interface }\label{app:1_3_data_appendix}
In this appendix, we expand on the analysis of the $\nu = \tfrac{1}{3}$/vacuum interface done in the main text. We present DMRG data for interactions beyond the simple $V_1$ pseudo-potential, such as a system with both $V_1,V_3$ interaction terms and screened Coulomb interaction. We also provide numerics from the composite fermion trial states, as described in Sec.~\ref{app:CF_trial}, which allow us to better interpret the DMRG results through the lens of optimizing the energy of a state with an excited composite fermion near the edge. 
We demonstrate that the composite fermion trial wavefunctions are in qualitative agreement with the DMRG data.

\subsection{Composite-fermion interpretation of the \texorpdfstring{$\nu = \tfrac{1}{3}$}{nu=1/3}/vacuum interface within the \texorpdfstring{$V_1$}{V1} interaction}
In this section, we analyse the results from the main text for the $\nu = \tfrac{1}{3}$/vacuum interface with the $V_1$ interaction, in the context of composite-fermion trial wavefunctions.

\begin{figure}[ht]
\includegraphics[width=0.4\linewidth]{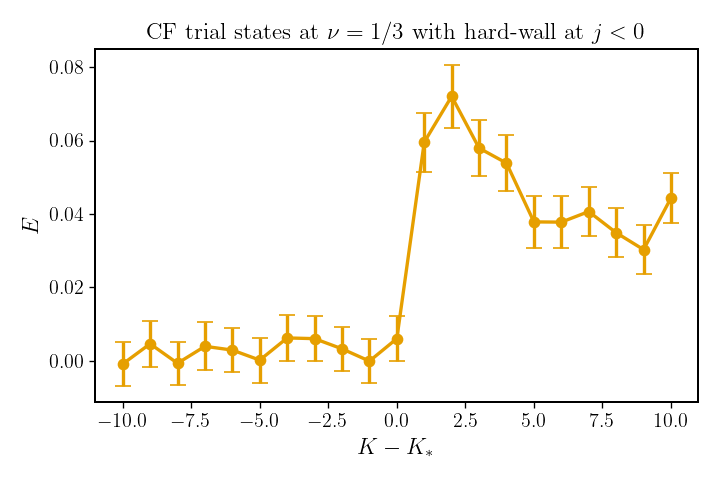}
\includegraphics[width=0.4\linewidth]{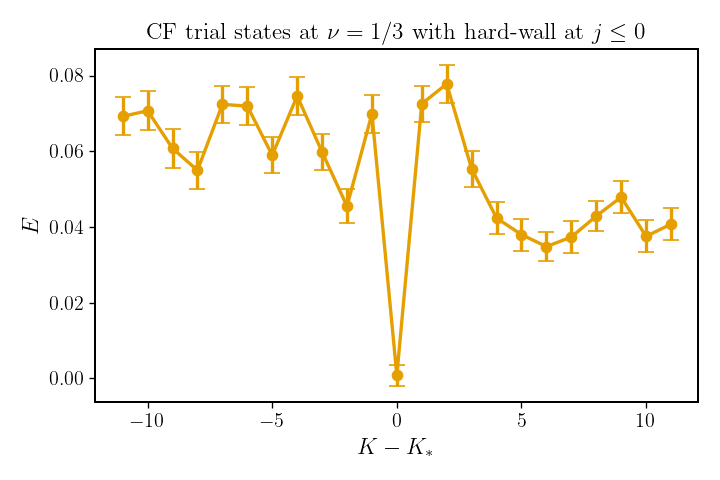}
         \caption{Illustration of a VMC calculation for \( \nu = \tfrac{1}{3} \) with the \( V_1 \) interaction: (a) with a hard wall at sites \( j < 0 \), where a quasihole can be moved into the bulk without any associated energy cost, and (b) with a hard wall at sites \( j \leq 0 \), where the intrinsic dipole is stabilized in the ground state.}
     \label{fig:monte_carlo_nu=1_3}
\end{figure}

In the case of a hard-wall confining potential at $j < 0$, for $K < K_{\PH}$ a composite fermion can occupy the $k = 0$ orbital without being promoted to a higher $\Lambda$ level, as discussed in Sec.~\ref{app:trial_CF_wavefunction_K_sectors}.
There is no large energy cost associated with moving a composite fermion from one momentum value to the other in the same $\Lambda$ level. Consequently, the $E(K)$ profile for $K \leq K_{\PH}$ is essentially flat.
For $K > K_{\PH}$, an electron is promoted from the ``reference'' state into the first $\Lambda$ level, which incurs an energy cost compared to $E(K_{\PH})$. 
As a result, the overall dispersion $E(K)$ is step-like, as can be seen in \autoref{fig:monte_carlo_nu=1_3}a.

In contrast, if a hard-wall confining potential is imposed at $j \leq 0$, the trial wavefunction for $K < K_{\PH}$ also requires promoting a composite fermion to a higher $\Lambda$ level from the “reference” state. This promotion is accompanied by a finite energy gap, thereby rendering the $K = K_{\PH}$ sector stable. The corresponding VMC calculation is shown in \autoref{fig:monte_carlo_nu=1_3}b. 
Ideally, in the future, we would like to perform an \( L_y \)-scaling analysis to confirm that the composite fermion trial wavefunctions agree with DMRG in the thermodynamic limit.

\subsection{\texorpdfstring{$V_1, V_3$}{V1,V3} interaction}
\begin{figure}[ht!]
    \centering
    \includegraphics[scale=0.5]{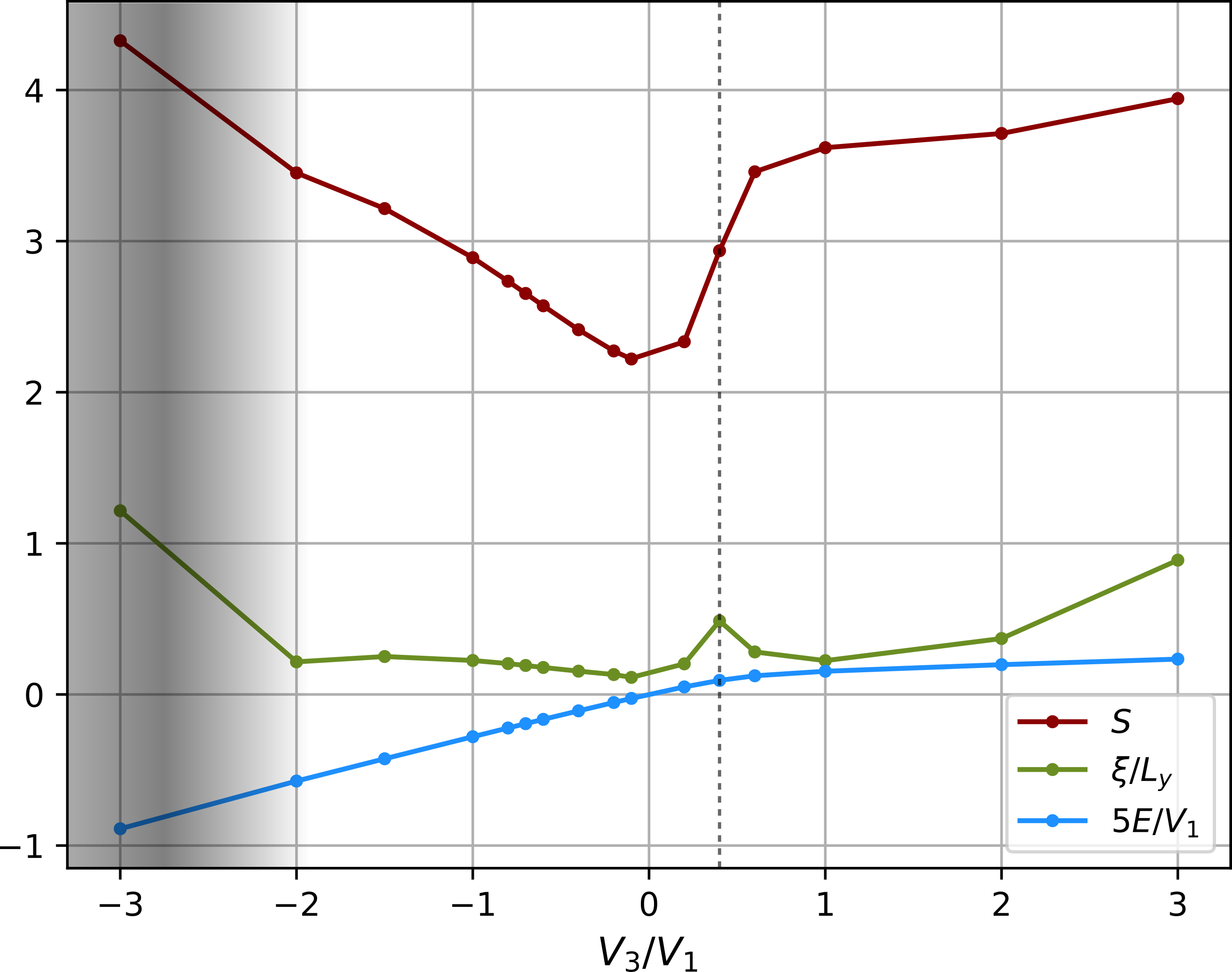}
    \caption{Plots of the Entropy $S$, the energy $E/V_1$, and the correlation length $\xi$ as a function of $V_3/V_1$. At $V_3/V_1\approx0.4$, there is a kink in the energy and a peak (might be finite or a proper divergence) of the correlation length, signalling a transition On the other side near $V_3/V_1 \approx -2$, the correlation length of the converged MPS state starts rapidly growing with decreasing $V_3$ and more involved calculations should be performed to assess the nature of the ground state in that regime.}
    \label{fig:idmrg_1_3}
\end{figure}

Going beyond a pure $V_1$ Haldane pseudopotential, we can consider interactions consisting of both $V_1$ and $V_3$ terms at filling factor $\nu = \frac{1}{3}$. By fixing $V_1=1$ as the energy scale of the system, we investigate how the ground-state dipole evolves by changing $V_3$. To study the effect $V_3$ has on the interface energetics, we should of course ensure that the $V_3$ term does not destabilize the Laughlin bulk ground state, which in turn fixes the intrinsic dipole. 
Starting from the exactly solvable point $V_3 = 0$, where the ground state is given by the Laughlin wavefunction, we vary $V_3$ and compute the corresponding bulk ground state via infinite DMRG, analysing the entanglement spectrum, entanglement entropy, energy, and correlation length, which are shown in \autoref{fig:idmrg_1_3}. Based on these results, we conclude that the state remains Laughlin for a range of potentials $-2\lesssim V_3\lesssim 0.4$, which is the range we will use when studying the interface dipole.

\FloatBarrier

\begin{figure}
\includegraphics[width=0.32\linewidth]{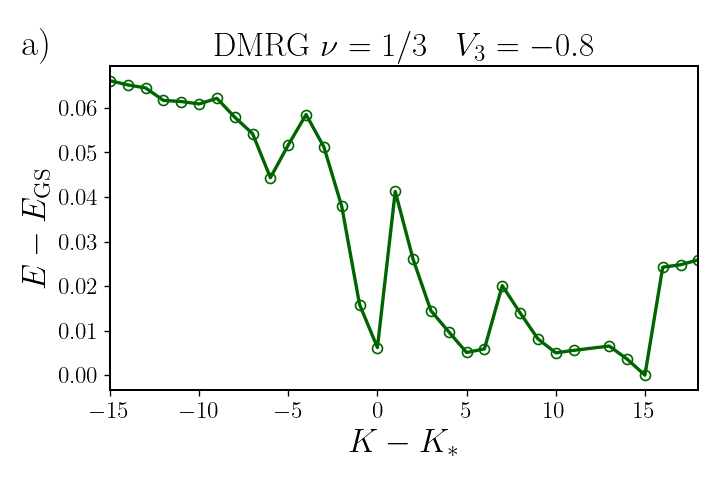}
\includegraphics[width=0.32\linewidth]{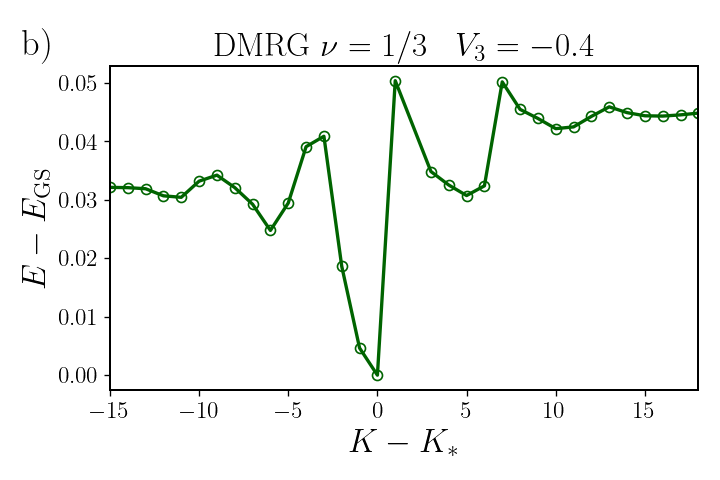}
\includegraphics[width=0.32\linewidth]{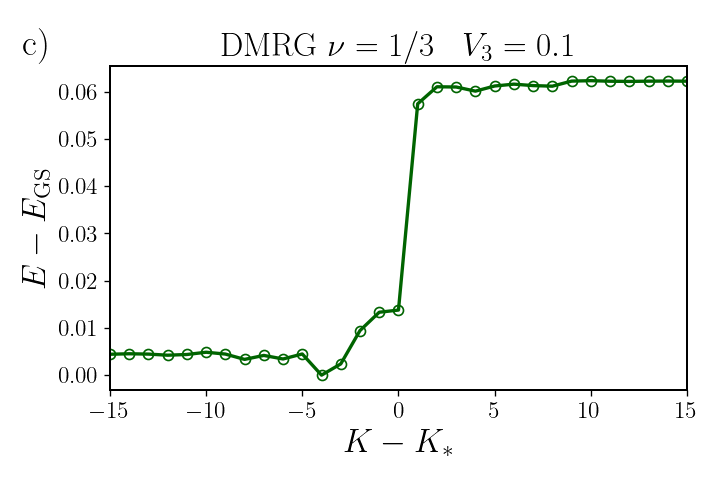}
\includegraphics[width=0.33\linewidth]{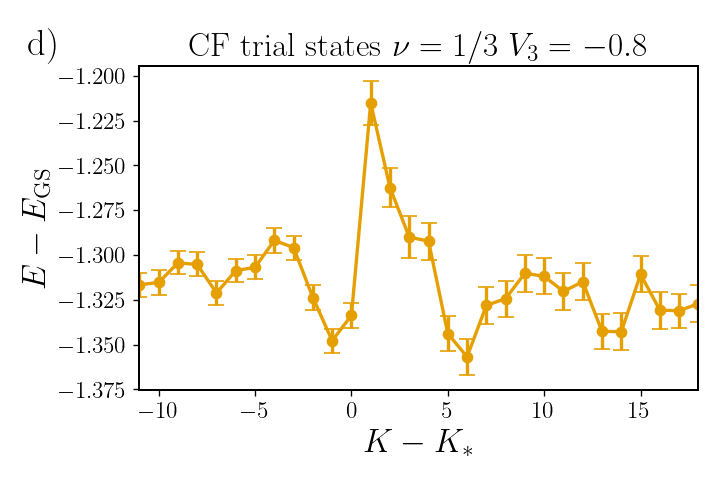}
\includegraphics[width=0.33\linewidth]{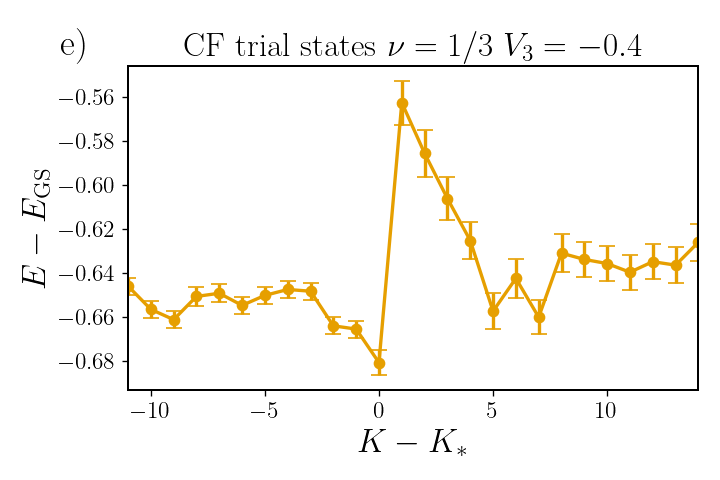}
     \caption{Subfigures (a)–(c) show the energy–\( K \) profiles for \( V_3 = -0.8, -0.4, 0.1 \), respectively. Subfigures (d) and (e) display the dispersions of the Monte Carlo trial wavefunctions for \( V_3 = -0.8 \) and \( V_3 = -0.4 \). The key observation is that, in panels (b) and (e), corresponding to \( V_3 = -0.8 \), the ground-state sector occurs at \( K \neq K_{\PH} \). All calculations are performed at \( L_y = 18\, \ell_B \).}

     \label{fig:DMRG:one third with V3}
\end{figure}

We now turn to an analysis of the dipole as the interaction parameter $V_3$ is varied, utilising both DMRG and VMC. The results are shown in \autoref{fig:DMRG:one third with V3}.
Throughout, we employ a hard-wall potential at $j < 0$. 
The results for a hard wall at $j \leq 0$ are qualitatively similar and are therefore not shown.
We begin by considering the regime $V_3 < 0$, starting from $V_3 = 0$ and decreasing $V_3$. For $0>V_3 \gtrsim -0.7$, the ground-state dipole coincides with the intrinsic dipole.
In this regime, we observe a discontinuous jump between $K(L_y) = K_{\PH}(L_y)$ and $K(L_y) = K_{\PH}(L_y) + 1$, corresponding to the energy cost of creating a quasielectron. 
There is no such jump in energy between $K(L_y) = K_{\PH}(L_y)$ and $K(L_y) = K_{\PH}(L_y) - 1$, but instead energy seems to increase smoothly.
Around $V_3 \approx -0.7$, the ground-state dipole undergoes a sharp change and shifts to a large positive value, $K (L_y)> K_{\PH}(L_y)$. 
The discontinuity in energy between $K(L_y) = K_{\PH}$ and $K(L_y) = K_{\PH}(L_y) + 1$ persists for $V_3<-0.7$. 
This indicates that the ground-state dipole no longer matches the intrinsic dipole. 
However, since the intrinsic dipole still corresponds to a local minimum of the energy, it may be possible to stabilise it by modifying the confining potential.

We complement our analysis using Monte Carlo composite fermion trial wavefunctions. These calculations predict a discontinuous energy jump on going from $K(L_y) = K_{\PH}(L_y)$ to $K(L_y) = K_{\PH}(L_y) + 1$. 
As $V_3$ is decreased, the composite fermion Monte Carlo results indicate that the ground state dipole undergoes a sudden shift from $K(L_y) = K_{\PH}(L_y)$ to a large positive $K$. This transition occurs around $V_3 \approx -0.7$, in quantitative agreement with the DMRG results.

We now turn to the regime $V_3 > 0$. In this case, the dipole value appears to be destabilized already for infinitesimally small values of $V_3$. 
This behaviour is illustrated in \autoref{fig:DMRG:one third with V3} c). The VMC using the composite fermion trial wavefunctions failed to capture this behaviour.

\FloatBarrier

\subsection{Coulomb interaction}

In this section, we present results for the $\nu=\frac{1}{3}$/vacuum interface with screened Coulomb interaction. 
We consider a hard wall at $j\leq 0$, in accordance with the definition given in the main text.

 \begin{figure}[ht]
  \includegraphics[width=0.32\linewidth]{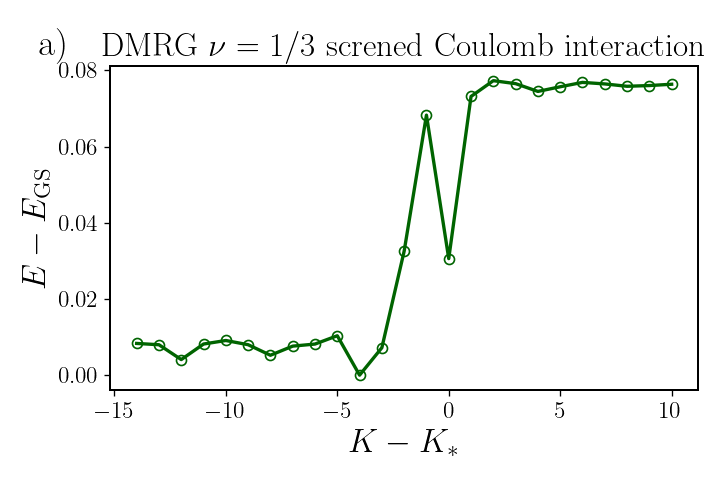}
\includegraphics[width=0.32\linewidth]{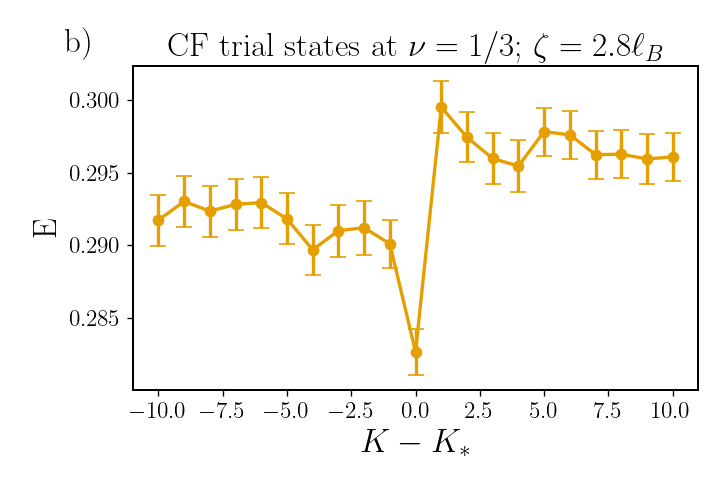}
\includegraphics[width=0.32\linewidth]{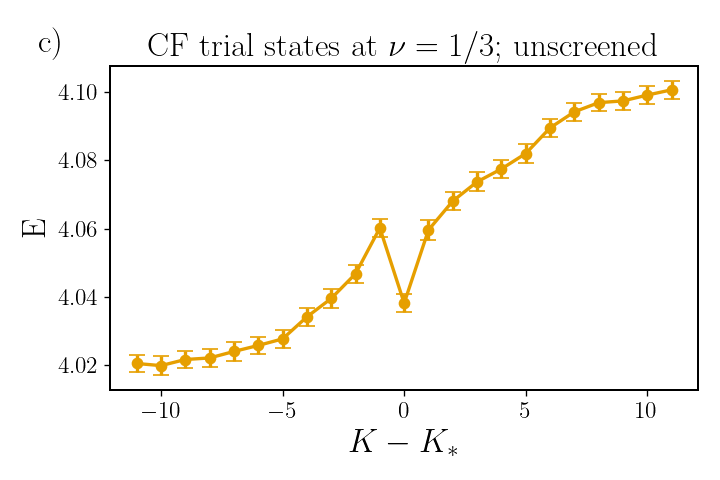}
     \caption{Figures showing the \( E(K) \) profile for the \( \nu=\frac{1}{3} \)/vacuum interface. Subfigure (a) presents DMRG results for a system with a hard wall at \( j \leq 0 \). Subfigure (b) shows the corresponding VMC results with a screened Coulomb interaction, while subfigure (c) shows the VMC results for the unscreened Coulomb interaction. For (a) and (b), the screening length is \( \zeta = 2.8\,\ell_B \). All data were obtained at \( L_y = 18.34\,\ell_B \).  } 
     \label{fig:Coulomb_1_3}
\end{figure}

For the hard wall at $j \leq 0$, the dipole at the global energy minimum appears to deviate from the intrinsic value.
However, the energy exhibits a discontinuous jump when moving from $K(L_y) = K_{\PH}(L_y)$ sector to $K(L_y) = K_{\PH} (L_y)\pm 1$ sector. 
This indicates that $K(L_y) = K_{\PH}(L_y)$ is locally stable, although it does not correspond to the global minimum. 
An appropriate choice of confining potential may therefore stabilize it as the global minimum.
A proper assessment would require finite-size scaling to determine whether the discontinuous jumps persist in the thermodynamic limit; however, such calculations are computationally demanding and were not performed here. The data is presented in \autoref{fig:Coulomb_1_3}.
The maximum bond dimension used to produce these figures is $\chi = 3000$, while the bulk ground state wavefunction was obtained with $\chi = 1500$.
Monte Carlo results for the hard wall at $j \leq 0$ also predict a cusp at $K = K_{\PH}$, indicating local stability. 
Monte Carlo simulations with a screened Coulomb interaction find that the cusp corresponds to a global minimum, whereas Monte Carlo simulations with the unscreened Coulomb interaction indicate that it is not a global minimum. 
Overall, the trial Monte Carlo results appears to be in qualitative agreement with the DMRG findings. 

\FloatBarrier

\section{\texorpdfstring{$\nu=\frac{2}{3}$}{nu=2/3}/vacuum interface}\label{app:2_3_data_appendix}
\subsection{Haldane \texorpdfstring{$V_1$}{V1} interaction --- finite size scaling}

In the main text, we concluded that the $\nu=\frac{2}{3}$/vacuum interface under the $V_1$ interaction with a hard-wall boundary does not exhibit an intrinsic dipole at the system size considered ($L_y=22\ell_B$).
As Eq.~\eqref{eq:HP_equation} holds only in the thermodynamic limit, to extend this conclusion, we have to examine the finite-size scaling with the circumference $L_y$. 
Since the dipole scales as
\begin{equation}
\frac{p^{x}}{L_y} \propto \frac{K(L_y)}{L_y^2},
\end{equation}
any state that does \emph{not} realize the intrinsic dipole in thermodynamic limit must have a momentum
\begin{equation} \label{eq:K_ref}
K_{\rm ref}(L_y) =K_{\PH}(L_y) -a L_y^2
\end{equation}
for some constant $a$. 

\begin{figure}[ht!]
\includegraphics[width=0.4\linewidth]{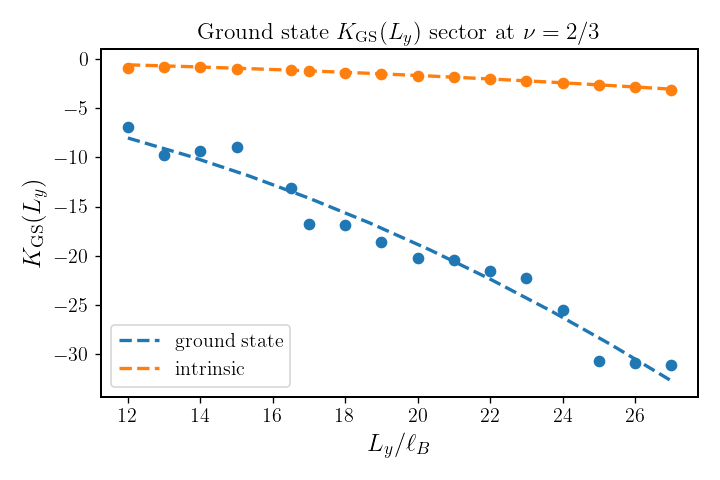}
     \caption{The figure compares the $L_y$ dependence of the ground-state momentum sector $K_{\mathrm{GS}}(L_y)$ with that of $K_{\PH}(L_y)$. We observe that $K_{\mathrm{GS}}(L_y)$ differs significantly from $K_{\PH}(L_y)$.}
     \label{fig:finite_Ly_scaling2}
\end{figure}

First, we aim to show that the ground state of the system obeys Eq.~\eqref{eq:K_ref}, which would directly imply that, in the thermodynamic limit, the ground state cannot attain the intrinsic dipole. To this end, we compare the ground-state sector $K_{\mathrm{GS}}(L_y)$ with $K_{\PH}(L_y)$ as a function of $L_y$, as presented in Eq. \eqref{fig:finite_Ly_scaling2}. We find that $K_{\mathrm{GS}}(L_y)$ differs from $K_{\PH}(L_y)$ at all system sizes considered. However, due to the limited clarity of the data, it is difficult to determine with absolute certainty whether $K_{\mathrm{GS}}(L_y) - K_{\PH}(L_y)$ follows a quadratic scaling of the form $a L_y^2$.

\begin{figure}
\includegraphics[width=0.4\linewidth]{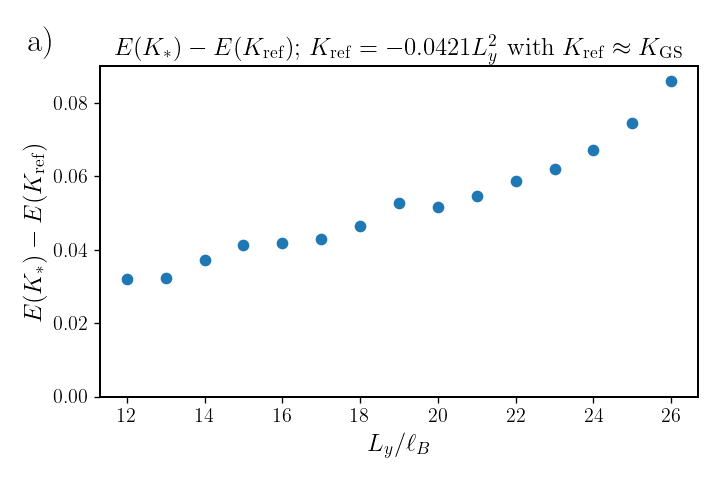}
\includegraphics[width=0.4\linewidth]{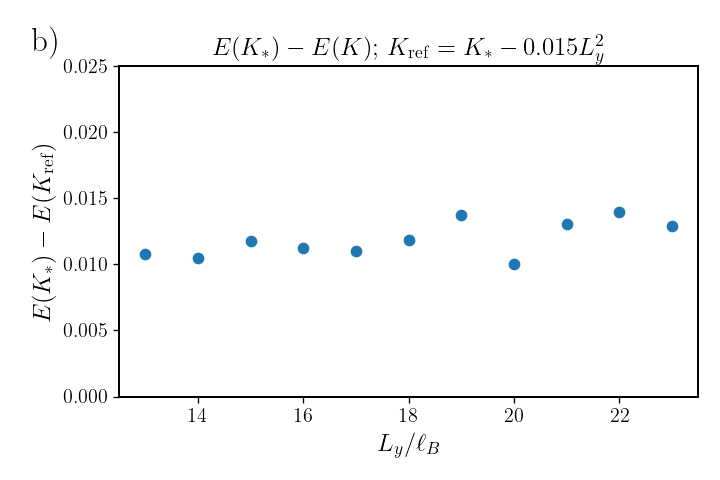}

     \caption{The energy difference $E(K) - E_{\PH}$, where $E_{\PH}$ is the energy at $K_{\PH}$, shown for a) $a$ chosen to approximately match $K_{\rm GS}$, and b) $a=0.015$.}
     \label{fig:finite_Ly_scaling}
\end{figure}

We further fit the momentum of the finite-size ground state to the scaling form $K_{\rm ref} = K_{\PH}(L_y) - a L_y^2$, and compute the corresponding energy difference $E(K_{\PH}) - E(K_{\rm ref})$. The results, shown in \autoref{fig:finite_Ly_scaling}a), indicate that, if this finite-size state is extrapolated to the thermodynamic limit, it retains dipole different than $p^{x}_{\PH}$ and has energy lower than that of the $K(L_y) = K_{\PH}(L_y)$ sector.

To provide more compelling evidence that the dipole cannot attain an intrinsic value in the thermodynamic limit, we fix a reference momentum of the form
$K_{\rm ref}(L_y) = K_{\PH}(L_y) - a L_y^2$, with a value of $a \approx 0.015$, and compare the corresponding energies. 
To reduce numerical noise, we evaluate $E(K_{\rm ref})$ by averaging over a small neighbourhood around each $K_{\rm ref}(L_y)$.

The key point is that, for the chosen value of $a$, the energy decreases monotonically as one moves from $K_{\PH}(L_y)$ to $K_{\rm ref}(L_y)$, so that no local minimum exists between these two points. 
This implies that, for each system size considered, there exists a state with dipole $p^x \neq p^x_{\PH}$ whose energy is lower than that of all states satisfying $|K - K_{\PH}| < a L_y^2$, provided that $E(K_{\PH}) - E(K_{\rm ref}) > 0$ persists in the thermodynamic limit.

We then verify this behaviour by plotting the energy difference $E_{\PH} - E(K_{\rm ref})$ as a function of $L_y$ in \autoref{fig:finite_Ly_scaling}b). For all system sizes considered, we find that $E(K_{\rm ref}) < E(K_{\PH})$. Moreover, the quantity \( E(K_{\mathrm{PH}}) - E(K_{\mathrm{ref}}) \) is non-decreasing as \( L_y \) increases, indicating that its sign is unlikely to change in the thermodynamic limit.

Consequently, if this behaviour persists as $L_y \to \infty$, the Park–Haldane dipole cannot be realized in the ground state. In particular, this rules out the possibility of a minimum occurring at deviations $|K - K_{\PH}| < b L_y$ for any constant $b$.

\FloatBarrier

\subsection{Coulomb interaction}

\begin{figure}[!htbp]
\includegraphics[width=0.32\linewidth]{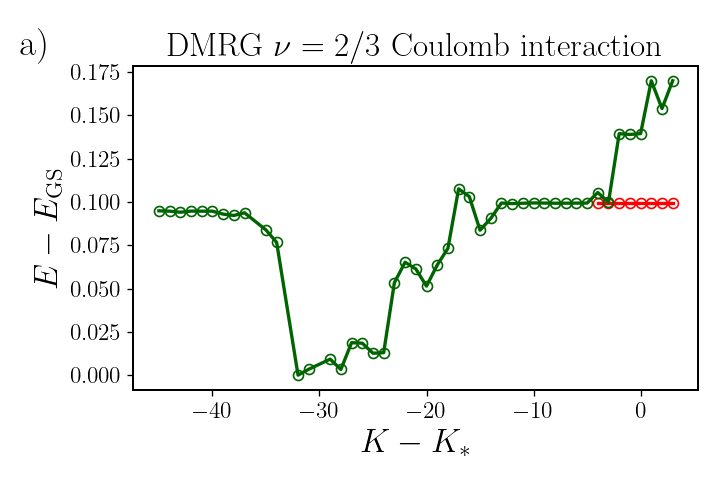}
\includegraphics[width=0.32\linewidth]{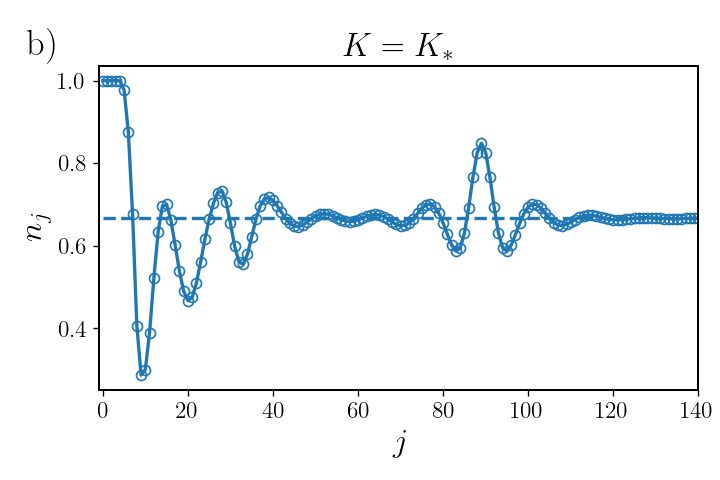}
\includegraphics[width=0.32\linewidth]{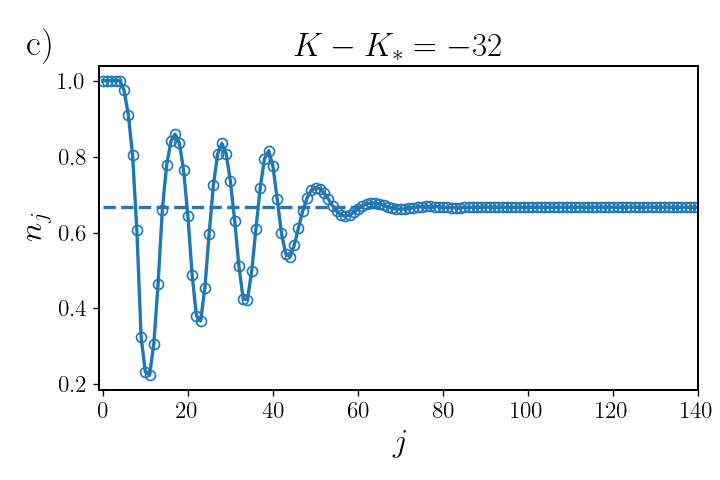}
 \caption{Subfigure a) shows the $E(K)$ dispersion for the $\nu = \tfrac{2}{3}$/vacuum interface with Coulomb interaction and perimeter $L_y = 19\,\ell_B$. The red curve represents our variational estimate based on moving  quasielectron.
Subfigure b) shows the density profile for the $K = K_{\PH}$ sector, while subfigure c) shows the density profile for the ground-state sector.
}
\label{fig:2_3_Coulomb_interaction}
\end{figure}

The data for $\nu=\frac{2}{3}$/vacuum interface with a screened Coulomb interaction in presence of a hard wall at $j< 0$ is remarkably similar to the $\nu=\frac{2}{3}$/vacuum data under the Haldane interaction presented in the main text (\autoref{fig:density_2_3}).
Here, we use a screening length of \( \zeta = 4 \ell_B \), and \( L_y = 19\, \ell_B \).
The DMRG data is presented in the \autoref{fig:2_3_Coulomb_interaction}.
We see that the $K(L_y)=K_{\PH}(L_y)$ state hosts a quasielectron in the bulk, at a position determined by the dipole value.
Hence, the $K(L_y)=K_{\PH}(L_y)$ is not the ground-state sector.
Changing the $K$ sector causes this quasielectron to move inwards or outwards in the bulk. 
Thus, the energy for $K>K_{\PH}$ remains constant.
The actual ground state occurs when the quasielectron is pushed completely to the edge at $K(L_y)-K_{\PH}(L_y)=-32$, corresponding to $p^{x} \approx -7.33$.
Unfortunately, we were unable to fully converge the data for $K(L_y) - K_{\PH}(L_y) > -2$ to the correct ground state. 
Instead, we present a trial variational ansatz obtained by taking the state at the $K(L_y) - K_{\PH}(L_y) > -10$ plateau and shifting the quasielectron outward.
This state has a lower energy than the wavefunction obtained via DMRG.

\section{Pfaffian/AntiPfaffian interface, additional data}\label{app:pf_apf_additional_data}
In this appendix, we present data for Pfaffian-AntiPfaffian interface for different values of $L_y$, $\zeta$, and topological sectors as an additional evidence that the ground state occurs at $p=0$, rather than at $p=p_{\PH}$. 

\begin{figure}[!htbp]

\includegraphics[width=0.4\linewidth]{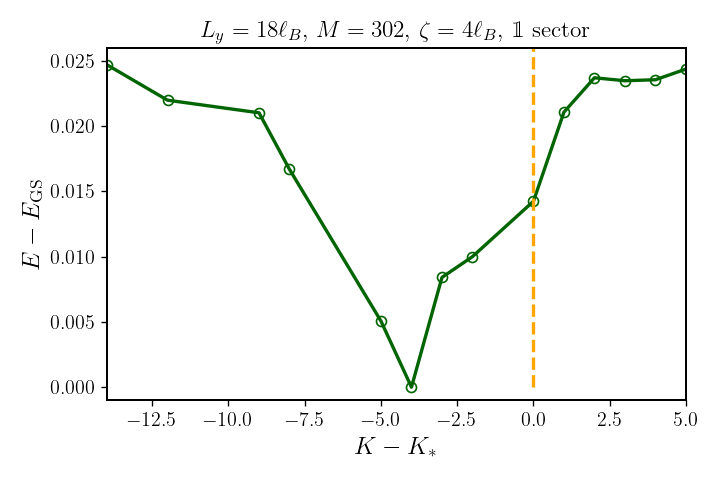}
\includegraphics[width=0.4\linewidth]{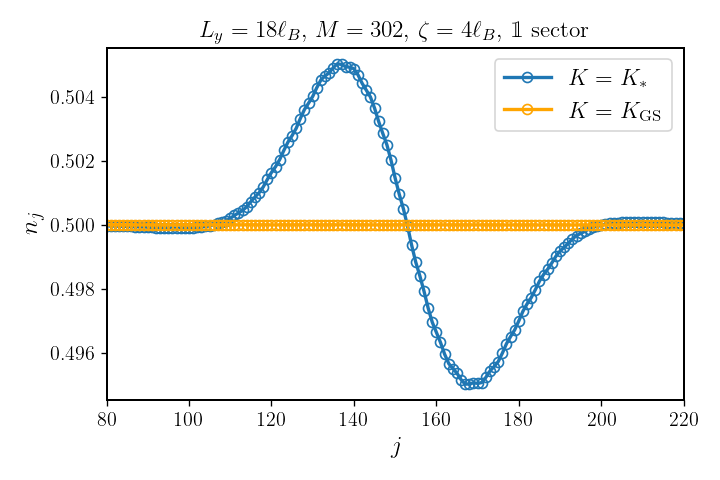}
\includegraphics[width=0.4\linewidth]{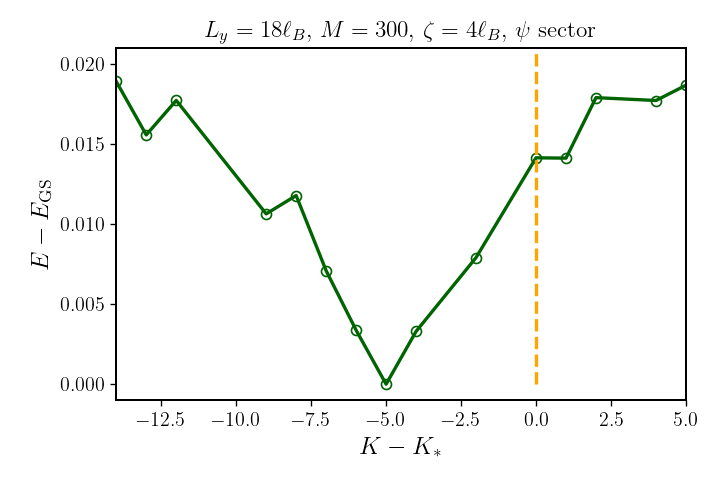}
\includegraphics[width=0.4\linewidth]{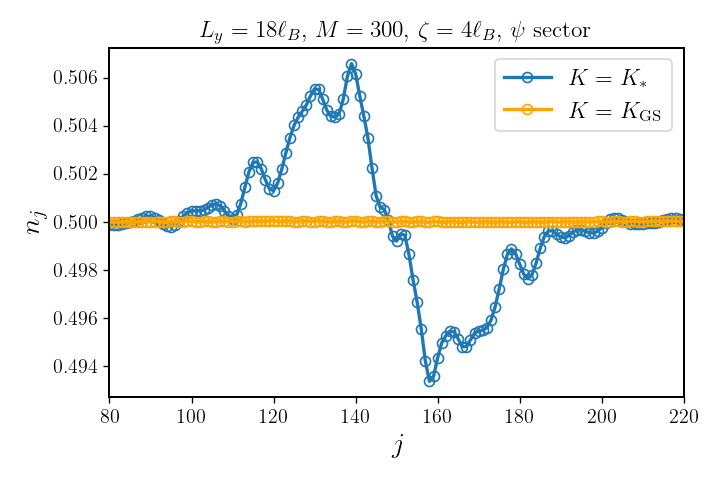}
\includegraphics[width=0.4\linewidth]{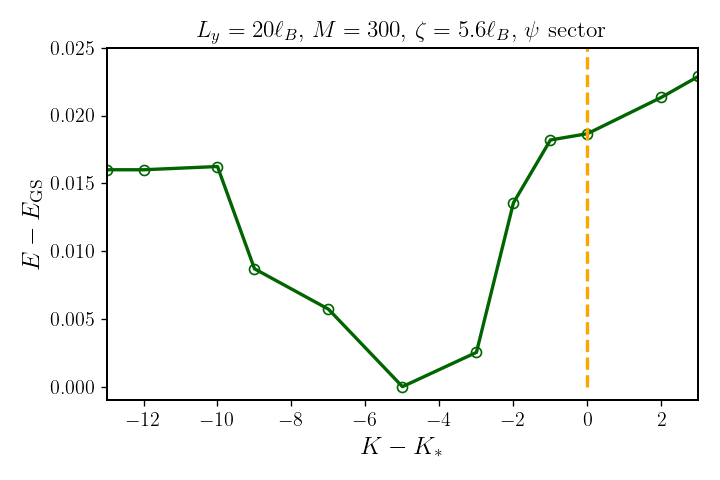}
\includegraphics[width=0.4\linewidth]{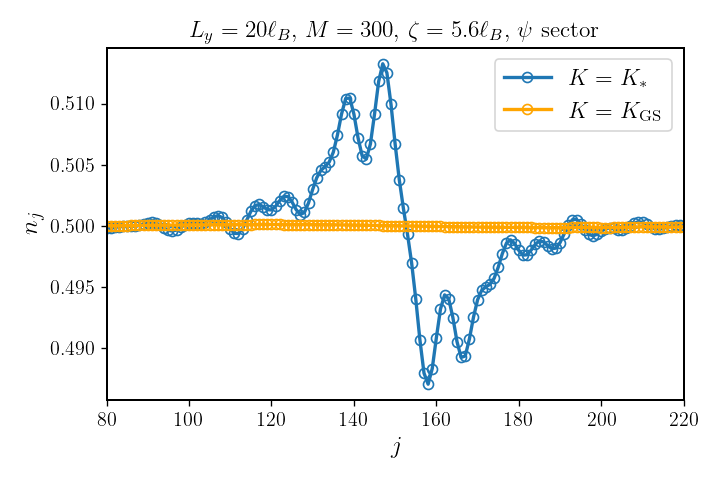}
 \caption{ Figures showing the \( E - K \) profiles for the Pfaffian–Anti-Pfaffian interface for different \( L_y \) and topological sectors (left). The figures on the right show the ground-state density profile and the \( K = K_{\PH} \) density profile. In all cases, the ground state corresponds to a flat density profile. }
\label{fig:pf_apf_additional data}
\end{figure}

We consider the case where the edge is in the topological identity sector $\mathbb{1}$, and the case where the edge is in $\psi$ sector. 
These correspond to segment lengths $M=302$ and $M=300$ respectively, for our particular choice of the environment.
Our conclusion that the ground state occurs for a flat density profile holds for both cases.
All other topological edge sectors break charge neutrality, and therefore the dipole cannot be examined for them. The data are shown in \autoref{fig:pf_apf_additional data}. The maximum bond dimension used for these calculations is $\chi=5000$, whereas the iDMRG is performed with $\chi=2600$. 

Finally, we note that achieving convergence across all $K$ sectors proved difficult, leaving open the possibility of a cusp somewhere; however, we consider this unlikely given the overall smoothness of the data.

\FloatBarrier

\section{Other filling fractions}\label{app:other_filling}
In this section of the Supplementary Information, we present data for additional filling fractions not included in the main text.

\subsection{\texorpdfstring{$\nu=\frac{2}{5}$}{nu=2/5}/vacuum interface}\label{app:2_5_data_appendix}

\subsubsection{Coulomb interaction}
\label{app:2_5_data_appendix_col}
We consider the data for $\nu=\frac{2}{5}$/vacuum interface with a screened Coulomb interaction in the presence of a hard wall $j\leq 0$.
Here, we use screening length $\zeta=4.2\ell_B$, and circumference $L_y = 19 \, \ell_B$. 
The DMRG data is presented in the \autoref{fig:2_3_Coulomb_interaction}.
We see that the $K(L_y)=K_{\PH}(L_y)$ state hosts a quasielectron in the bulk, and that $K(L_y)=K_{\PH}(L_y)$  is not the ground-state sector.
The actual ground state occurs when the quasielectron is pushed completely to the edge at $K(L_y)-K_{\PH}(L_y)=-20$.

\begin{figure}
    \includegraphics[width=0.32\linewidth]{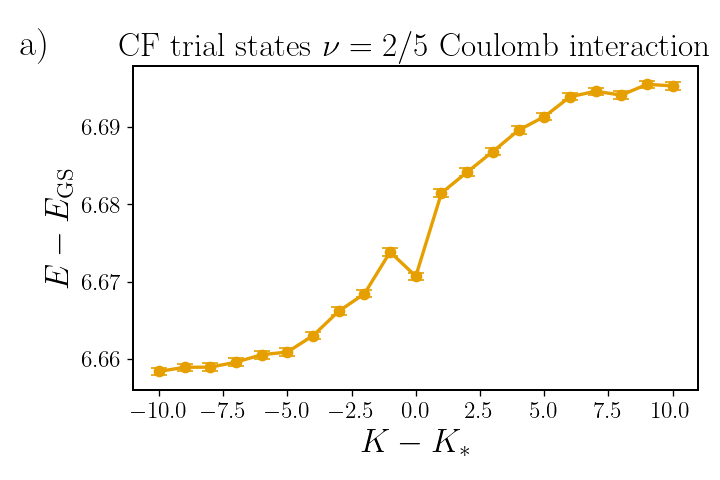}
    \includegraphics[width=0.32\linewidth]{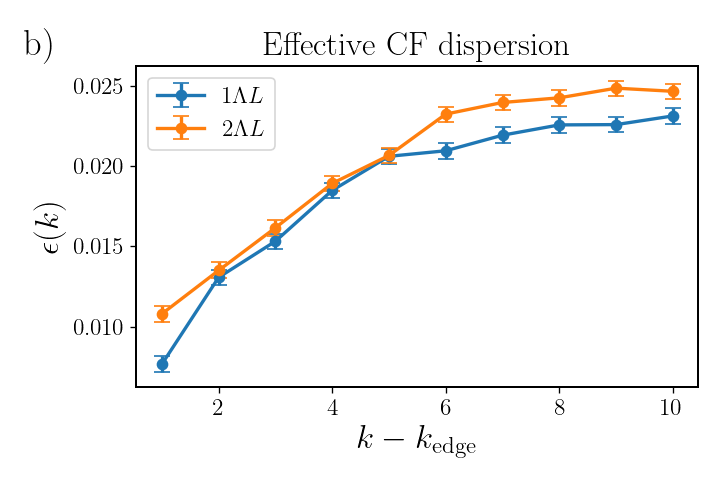}
    \caption{ Variational Monte Carlo results for trial composite fermion states at $\nu=\frac{2}{5}$ with unscreened Coulomb interaction. (a) Energy as a function of $K-K_{\PH}$. (b) Extracted effective composite fermion dispersion for the $1\Lambda$ and $2\Lambda$ levels, illustrating their relative energetics near the edge. The energy levels plateau to a constant deep in the bulk. }
\label{fig:MC_2_5_Coulomb_interaction}
\end{figure}

We also analyse trial composite fermion wavefunctions.
For small screening lengths, VMC finds that the \( K = K_{\PH} \) state is the global minimum.
For larger screening lengths, however, this state remains only a local minimum and is no longer globally stable.
We therefore focus on the large screening length (effectively unscreened) regime, as it yields results consistent with DMRG.
The origin of the discrepancy in the location of the global minimum between VMC and DMRG is not fully understood and is left for future work.
The data are presented in \autoref{fig:MC_2_5_Coulomb_interaction}.

As noted earlier, within the composite fermion picture, the state with $K(L_y)=K_{\PH}(L_y)$ is not the global energy minimum.
Instead, the global minimum appears at $K(L_y)<K_{\PH}(L_y)$.
We do not expect trial wavefunctions obtained by moving a \emph{single} composite fermion to accurately capture the physics of $\nu=\frac{2}{5}$.
Consistent with this, the DMRG density profile shows an enhanced occupancy near the edge, whereas this feature is absent in the VMC results for our trial wavefunctions.
The VMC energy profile $E(K)$ is shown in \autoref{fig:MC_2_5_Coulomb_interaction}, from which we can draw further conclusions about the energetics.

We provide an intuition using the effective dispersion of composite fermions.
We label by $k_{\rm edge}=1$ the leftmost occupied CF orbital momentum number.
In the case of hard wall at $j\leq 0$ , we are not allowed to fill $\Lambda$ levels past $k<k_{\rm edge}$.
Thus, if one expects effective bands to simply curve up near the edge as in \autoref{fig:cf_dispersion}, one would naively expect the ground state to possess an intrinsic dipole.
However, if the minimum energy of the $2\Lambda$ level lies below the maximum energy of the $1\Lambda$ level, the intrinsic dipole is expected to be destroyed.

\begin{figure}
\includegraphics[width=0.32\linewidth]{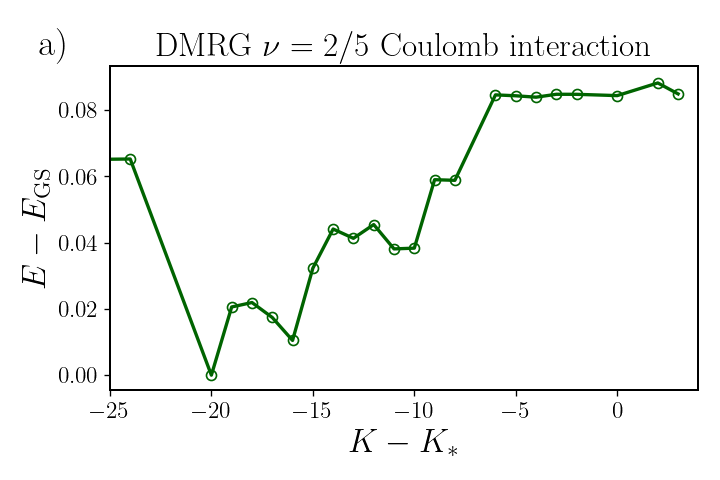}
\includegraphics[width=0.32\linewidth]{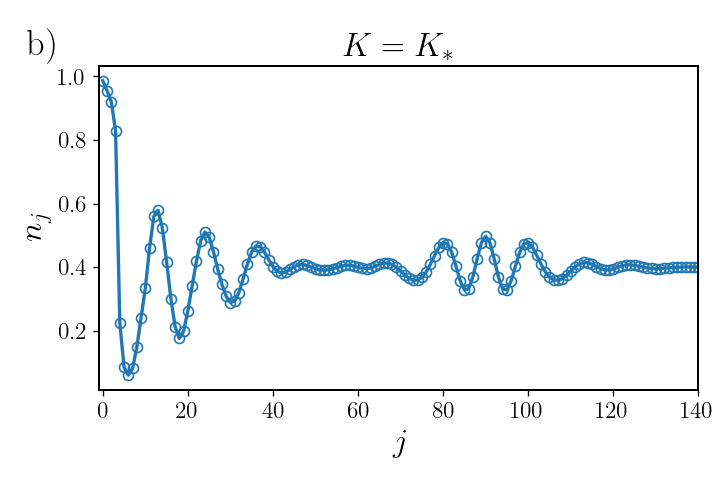}
\includegraphics[width=0.32\linewidth]{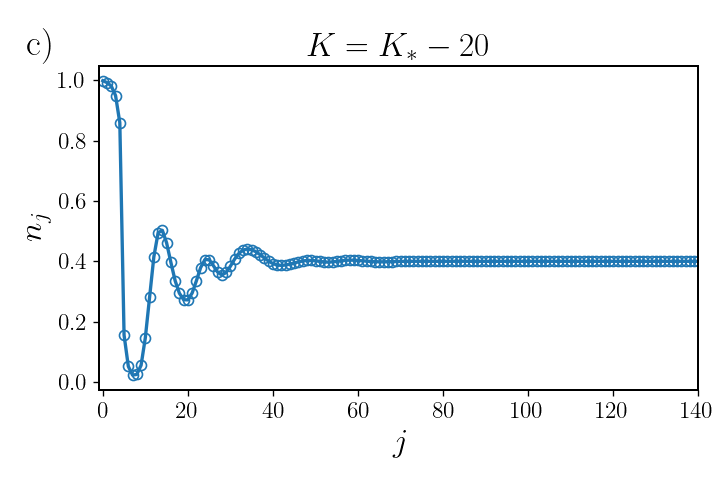}
 \caption{DMRG results for the $\nu=\frac{2}{5}$/vacuum interface with screened Coulomb interaction ($\zeta=4.2\ell_B$, $L_y=19\ell_B$) and a hard wall at $j\leq 0$. Subfigure (a) shows the energy $E-E_{\min}$ versus momentum shift $K-K_{\PH}$. Subfigures (b) and (c) show the orbital occupations \( n_j \) for \( K = K_{\mathrm{PH}} \) and \( K = K_{\mathrm{PH}} - 20 \), respectively, illustrating the transfer of the quasielectron from the bulk to the edge.}
    \label{fig:2_5_Coulomb_interaction}
\end{figure}

Using the energetics in \autoref{fig:MC_2_5_Coulomb_interaction}a), we can derive the effective composite fermion dispersion for $1\Lambda$ level and $2\Lambda$ level, depicted in \autoref{fig:MC_2_5_Coulomb_interaction}b).
We find that an effective dispersion has  $2\Lambda$ level near $k=k_{\rm edge}$ lower in energy than the $1\Lambda$ level near $k=k_{\rm edge}+5$, thus destroying the intrinsic value of the dipole. 
 We expect three $\Lambda$ levels to be filled near the edge, leading to a build-up of density in that region, consistent with the density profiles predicted by DMRG.

Furthermore, the composite-fermion approach suggests that, in the ground state, several quasielectron–quasihole pairs will be present near the edge. 
To capture the ground-state wavefunction more accurately, it would be beneficial to combine exact diagonalization (ED) with variational Monte Carlo (VMC). 
In particular, one could construct an effective Hamiltonian that allows for up to $N_0$ quasielectron–quasihole pairs and then diagonalize it.
Naturally, the resulting effective dispersion can depend sensitively on the specific configuration of composite fermions under consideration. 
A more detailed study will be presented in future work.

\subsubsection{\texorpdfstring{$V_1=1,V_3=0.05$}{V1} interaction}

\begin{figure}[!htbp]
\includegraphics[width=0.45\linewidth]{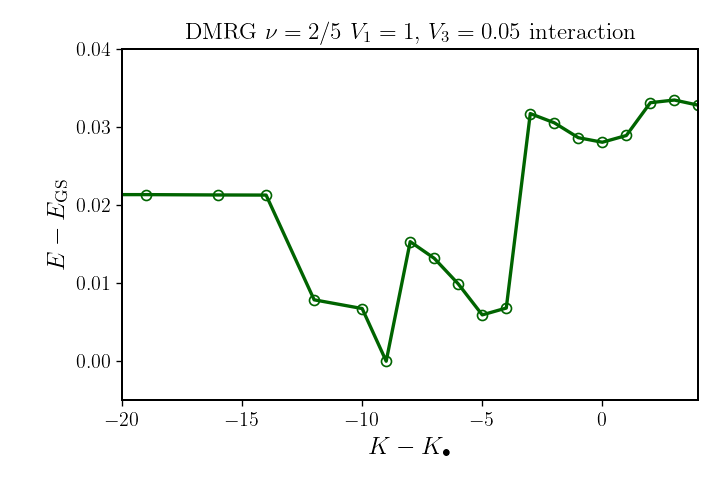}

 \caption{Figure shows the dispersion for $\nu=\frac{2}{5}$/vacuum boundary within the interaction set by $V_1=1$, $V_3=0.05$ in the hard-wall confining potential. Here we use $L_y=18\ell_B$, and maximum bond dimension $\chi=2500$, $M=176$.}
\label{fig:2_5_Haldane_interaction}
\end{figure}

We provide the results for $\nu=\frac{2}{5}$/vacuum in the pseudopotential interaction set by $V_1=1$ and $V_3=0.05$ pseudopotentials, with hard-wall confining potential at $j<0$.
DMRG results are shown in the \autoref{fig:2_5_Haldane_interaction}.
It can be seen that the dipole in the ground state does not take an intrinsic value.
Interestingly, the density profile does not exhibit a density pile-up, in contrast to the case of the Coulomb interaction presented in sec. \ref{app:2_5_data_appendix_col}.

In the variational Monte Carlo calculation, there is no sharp energy gap observed when exciting a single composite fermion from \( k = k_{\rm edge} \) in the \( 1\Lambda \) level to \( k = k_{\rm edge} - 1 \) in the \( 2\Lambda \) level - unlike what is seen with Coulomb interactions. This suggests that creating a quasielectron-quasihole pair near the edge incurs no significant energy cost.
As a result, to accurately determine composite fermion occupancies, it is necessary to consider an effective Hamiltonian that includes contributions from many quasielectron--quasihole pairs.

\FloatBarrier

\subsection{\texorpdfstring{$\nu=\frac{1}{5}$}{nu=1/5}/vacuum interface}
\begin{figure}[!htbp]
  \includegraphics[width=0.35\linewidth]{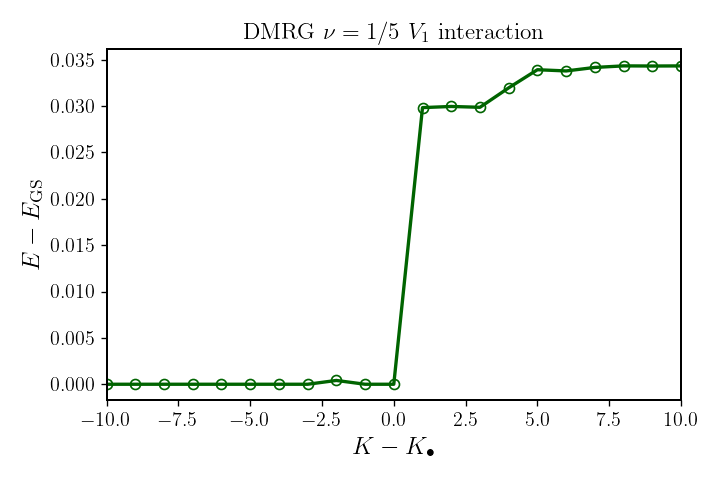}
  \includegraphics[width=0.35\linewidth]{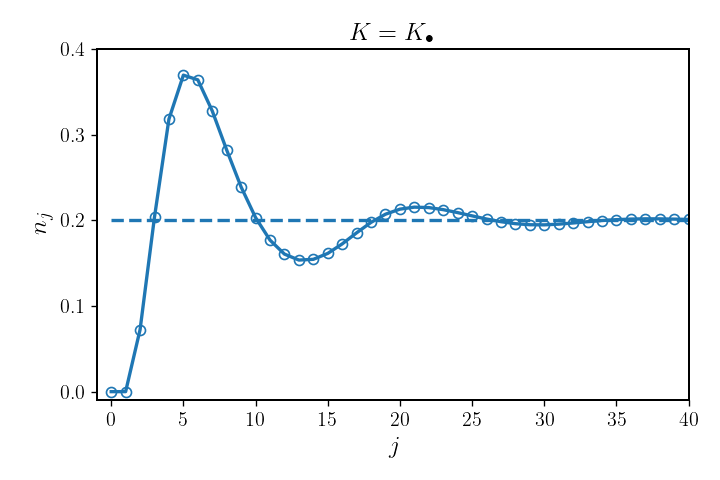}
     \caption{In this figure we show the $\nu=\frac{1}{5}$/vacuum boundary for the $V_1=1,V_3=1$ interaction with hard-wall confining potential. Figure a) represents the energy-$K$ profile, showing that $K<K_{\PH}$ states have the same energy as the state at $K=K_{\PH}$. Figure b) shows the density profile of the $K=K_{\PH}$ sector. We can see that sites labelled by $j=0,1$ are empty. 
     For this calculation we use $L_y=18\ell_B$.}
     \label{fig:DMRG_one_fifth}
\end{figure}

In this section, we study the interface between the vacuum and a $\nu=\frac{1}{5}$ state for an interaction defined by the pseudopotentials $V_1 = 1$ and $V_3 = 1$. 
The confining potential used is a hard wall at $j < 0$. 
Hamiltonian defined by the $V_1$ and $V_3$ pseudopotentials is a parent Hamiltonian of the Laughlin $\nu=\frac{1}{5}$ state. 
We therefore expect behavior similar to the $\nu=\frac{1}{3}$ case with $V_1 = 1$.
The DMRG data is presented in \autoref{fig:DMRG_one_fifth}. 
From the density profile $n_j$ at $K(L_y) = K_{\PH}(L_y)$, one observes that the first two sites are empty, in contrast to the single empty site at the boundary of $\nu=\frac{1}{3}$ Laughlin state. 
This can be explained by counting powers of \( w_i \) in the Laughlin wavefunction, as discussed in Sec.~\ref{app:CF_trial}.
Consequently, to make $K(L_y) = K_{\PH}(L_y)$ a stable ground state, the position of the confining wall must be shifted inward by two sites, rather than by one as for $\nu=\frac{1}{3}$/vacuum interface. 
In other words, to stabilise the \( K(L_y) = K_{\PH}(L_y) \) state, one should impose a hard wall at \( j \leq 1 \).

\FloatBarrier

\subsection{\texorpdfstring{$\nu=\frac{3}{7}$}{nu=3/7}/vacuum interface }

In this section, we examine $\nu=\frac{3}{7}$/vacuum interface in the presence of a hard-wall potential at $j \leq 0$, with screened Coulomb interaction. The parameters used were $L_y=22\ell_B$ and $\zeta=3\ell_B$.

To verify that the $\nu = \frac{3}{7}$ state obtained from iDMRG lies in the same topological phase as the corresponding higher hierarchy state, we compute its topological shift $\mathcal{S}$. 
The shift is extracted from the expectation value $\langle K \rangle$ following the procedure of \cite{Zaletel_2013}. 
The resulting value, $\mathcal{S} = 5$, is in agreement with the theoretical prediction.
We notice that shift directly dictates the value of dipole \cite{HaldanePark}.

\begin{figure}[ht]
\includegraphics[width=0.32\linewidth]{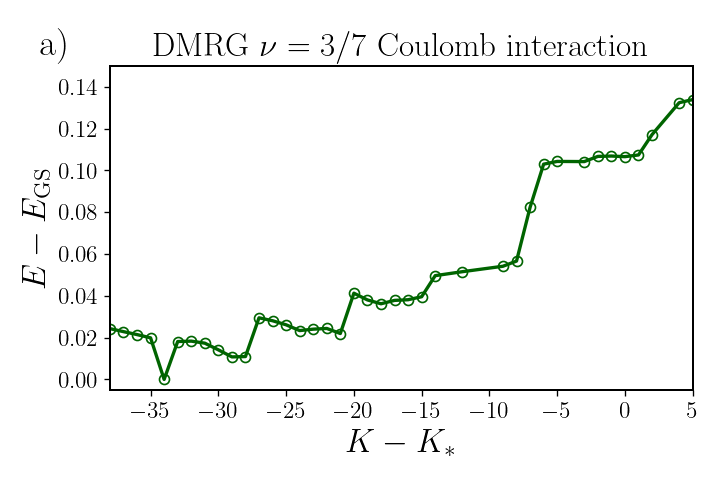}
\includegraphics[width=0.32\linewidth]{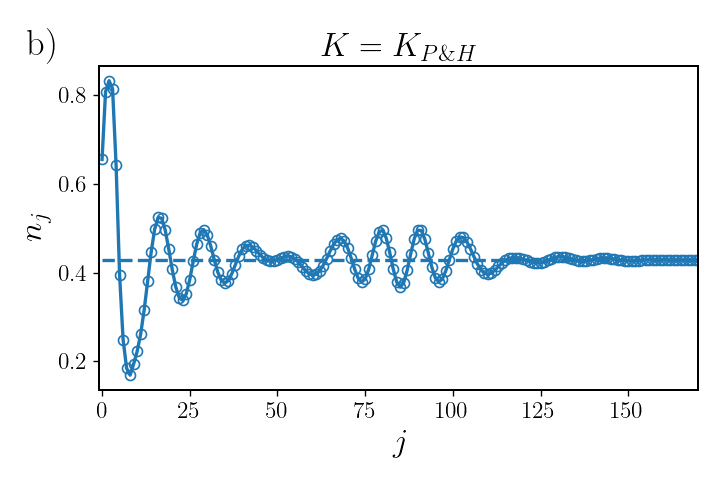}
\includegraphics[width=0.32\linewidth]{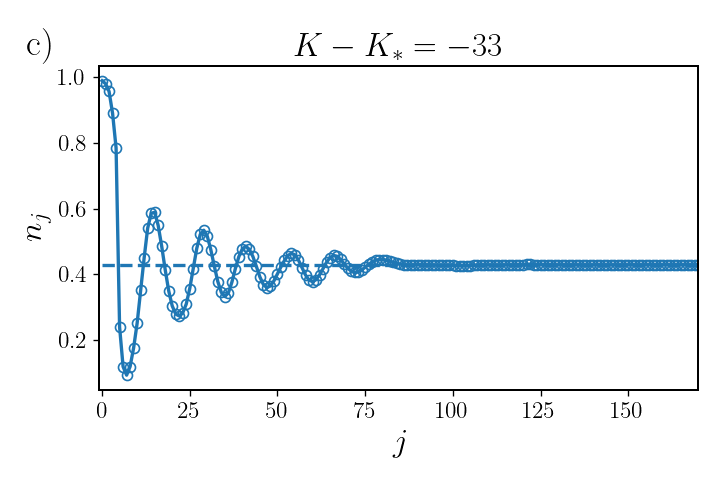}

 \caption{The data for the \( \nu = \tfrac{3}{7} \)/vacuum interface is presented. Subfigure (a) shows the \( E(K) \) dependence. Subfigure (b) shows the density profile of the \( K = K_{\mathrm{PH}} \) sector, while subfigure (c) shows the density profile of the ground-state sector. Here we use \( L_y = 22\, \ell_B \), \( M = 176 \), and \( \zeta = 3\, \ell_B \).}
     \label{fig:3_7_dmrg_data}
\end{figure}

The DMRG data for $\nu=\frac{3}{7}$ is shown in \autoref{fig:3_7_dmrg_data}. For the $\nu=\frac{3}{7}$/vacuum interface, we find that the ground state lies far from the $K = K_{\PH}$ sector.
Density profile at $K=K_{\PH}$ contains a lone quasielectron of charge $1/7$.
For $K > K_{\PH}$, this quasielectron is pushed outward toward the bulk, with a small energy cost. In contrast, in the ground state sector, the density profile does not show any clear signature of an isolated quasielectron or quasihole.

\FloatBarrier
\end{document}